\documentclass[reprint,
superscriptaddress,
nobibnotes,
amsmath,amssymb,
aps,
longbibliography
]{revtex4-2}
\usepackage[normalem]{ulem}
\usepackage{dcolumn}
\usepackage{amssymb, esint}
\usepackage{physics}
\usepackage{bm}
\usepackage{graphicx}
\usepackage[colorlinks,allcolors=blue]{hyperref}
\usepackage{url}
\usepackage{verbatim}

\def\be{\begin{equation}}   
\def\ee{\end{equation}}
\def\bfr{{\bf r}}   
   
\def\bfk{{\bf k}}   
\def\bfq{{\bf q}} 
\def\bfR{{\bf R}}   
\def\bfA{{\bf A}} 
\def\bfP{{\bf P}}

\def\bfG{{\bf G}}

\def\bfM{{\bf M}}

\def\bfn{{\bf n}}   
\def\bfp{{\bf p}}

\newcommand{\bnabla}{\boldsymbol{\nabla}}
\newcommand{\bAh}{\mathbf{\hat{A}}}
\newcommand{\beps}{\boldsymbol{\epsilon}}
\newcommand{\imagi}{\mathrm{i}}

\def\ket#1{| #1 \rangle}

\usepackage{color}

\usepackage[normalem]{ulem}

\usepackage[nolist]{acronym}

\begin{document}

\title{Effective Equilibrium Theory of Quantum Light-Matter Interaction in Cavities:\newline
Extended Systems and the Long Wavelength Approximation}

\author{Mark Kamper Svendsen}
\email{mark.svendsen@nbi.ku.dk}
\affiliation{Max Planck Institute for the Structure and Dynamics of Matter and Center for Free-Electron Laser Science, Luruper Chaussee 149, 22761, Hamburg, Germany}

\author{Michael Ruggenthaler}
\affiliation{Max Planck Institute for the Structure and Dynamics of Matter and Center for Free-Electron Laser Science, Luruper Chaussee 149, 22761, Hamburg, Germany}

\author{Hannes Hübener}
\affiliation{Max Planck Institute for the Structure and Dynamics of Matter and Center for Free-Electron Laser Science, Luruper Chaussee 149, 22761, Hamburg, Germany}

\author{Christian Schäfer}
\affiliation{Department of Microtechnology and Nanoscience, MC2, Chalmers University of Technology, 412 96 Göteborg, Sweden}
\affiliation{Department of Physics, Chalmers University of Technology, 412 96 Göteborg, Sweden}

\author{Martin Eckstein}
\affiliation{Department of Physics, University Hamburg, D-22607 Hamburg, Germany}

\author{Angel Rubio}
\email{angel.rubio@mpsd.mpg.de}
\affiliation{Max Planck Institute for the Structure and Dynamics of Matter and Center for Free-Electron Laser Science, Luruper Chaussee 149, 22761, Hamburg, Germany}
\affiliation{Initiative for Computational Catalysis (ICC), The Flatiron Institute, 162 Fifth avenue, New York NY 10010, United States}

\author{Simone Latini}
\email{simola@dtu.dk}
\affiliation{2DPHYS, Department of Physics, Technical University of Denmark, 2800 Kgs.~Lyngby, Denmark}
\affiliation{Max Planck Institute for the Structure and Dynamics of Matter and Center for Free-Electron Laser Science, Luruper Chaussee 149, 22761, Hamburg, Germany}

\begin{abstract}
When light and matter interact strongly, the resulting hybrid system inherits properties from both constituents, allowing one to modify material behavior by engineering the surrounding electromagnetic environment. This concept underlies the emerging paradigm of cavity materials engineering, which aims at the control of material properties via tailored vacuum fluctuations of dark photonic environments. The theoretical description of such systems is challenging due to the combined complexity of extended electronic states and quantum electromagnetic fields. Here, we derive an effective, non-perturbative theory for low-dimensional crystals embedded in a Fabry-Pérot resonator, within the long-wavelength limit. Our approach incorporates the multimode and dispersive nature of the cavity field and reduces it to an effective single-mode description by imposing the condition of negligible momentum transfer from light to matter. Importantly, the resulting effective mode is characterized by a finite mode volume—even in the limit of extended cavities—which is directly linked to realistic cavity parameters. This ensures that the light–matter coupling remains finite in bulk systems. By explicitly accounting for the finite reflectivity of cavity mirrors, our theory also avoids double counting the contribution from free-space light–matter coupling. Overall, our results provide a robust and realistic framework for describing cavity–matter interactions at the Hamiltonian level, incorporating the electromagnetic environment beyond the idealized perfect-mirror approximation.
\end{abstract}

\maketitle

\begin{acronym}
\acro{LWA}[LWA]{long-wavelength approximation}
\end{acronym}

\section{Introduction}\label{sec:intro_sec}

Historically, the physics of strong light-matter interactions has primarily drawn interest in the field of quantum optics~\cite{carusotto2013quantum}. However, in recent years the possibility of exploiting strong-light matter coupling for material design and tuning of chemical properties has sparked growing interest in the condensed-matter community. When matter interacts strongly with light, hybrid light-matter states, called \textit{polaritons}, emerge. Because these polaritonic states inherit properties from both constituents, it is possible to alter the properties of the coupled system by changing either of the two~\cite{Basov2020, bloch2022strongly, hubener2024quantum}.  Strong light-matter coupling~\cite{forn2019ultrastrong} has for example been exploited to alter the optical properties in semi-conductors and in quantum Hall systems~\cite{hagenmuller2010ultrastrong,Smolka2014,todorov20152DEG,Bagliani2018,Ravets2018,knuppel2019nonlinear,rokaj2019quantum,Cortese20192DEG,cortese2021excitons,rokaj2022free,Keller2020,rokaj2022polaritonic,appugliese2022breakdown,de2022magnetic,Ciuti2023transport}. More recently, the appealing scenario of modifying equilibrium material properties via the coupling to the electromagnetic fluctations of a dark dielectric environment has led to the proposal of light-mediated superconductivity originating from cavity induced electron-pairing~\cite{Schlawin2019} or from polariton condensation~\cite{cotlect2016superconductivity}. Tunable superconductive temperature has also been proposed for both a FeSe/SrTiO$_3$~\cite{Sentef2018} system and MgB$_2$~\cite{lu2024cavity} bulk crystal, as a result of the polaritonic enhancement of the electron-phonon coupling, and in the case of MgB$_2$ the critical superconductive temperature is increased. In general, it is worth mentioning that different point of views have been proposed in the literature on the possibility of mediating quantum phase transitions through cavities and the reader should consult further literature~\cite{andolina2024amperean,Mazza2019,Andolina2019}. There is, however, agreement on the fact that in the presence of material non-linearities and/or multiple photonic modes such transitions cannot be excluded~\cite{Andolina2020, Vidal2021, bloch2022strongly, hubener2024quantum}. In a recent work~\cite{Huebener2021}, the idea of using circularly polarized cavities to alter the topology of a crystal has been put forward and theoretically demonstrated by the appearance of a quantized Hall conductance associated with an integer Chern number in graphene~\cite{Wang2019, dag2024engineering}. Strong coupling of Tera-Hertz (THz) cavities to the ferroelectric soft phonon mode of SrTiO$_3$ has been proposed to aid the paraelectric-to-ferroelectric phase transition~\cite{Ashida2020} and generate a ferroelectric photo-groundstate~\cite{Latini2021-2}. A recent experiment reported a 50 K reduction in the transition temperature for the metal-to-insulator phase transition in 1T-TaS$_2$ when the TaS$_2$ crystal is embedded in a Giga-Hertz (GHz) cavity~\cite{jarc2023cavity}. Similar temperature effects have been proposed in the context of strong light-matter coupling of molecular systems~\cite{sidler2022perspective,sidler2022exact,ruggenthaler2022understanding}. Several other interesting predictions and effects are discussed in recent reviews~\cite{schlawin2022cavity, ebbesen2023introduction, lu2025cavity}.

The set of methods employed in theoretical work on cavity-matter systems has been scattered and often developed from a different underlying theory. For molecular systems, efforts have been devoted to generalize quantum chemistry methods, such as configuration interaction and coupled cluster, to quantum electrodynamics (QED) to provide a numerical approach for accurate calculations of molecular properties in idealized electromagnetic environments~\cite{Haugland2020, mordovina2020polaritonic, Haugland2021, liebenthal2022equation, foley2023ab, riso2023strong, romanelli2023effective, pavosevic2023cavity, datta2024coupled, vu2024cavity}. The density functional theory generalization to quantum electrodynamics has initially been made practical for the prediction of groundstate properties and the linear response of molecular systems in idealized cavities~\cite{ruggenthaler2014quantum, pellegrini2015optimized,flick2018ab, ruggenthaler2018quantum, flick2019light,schafer2021making,flick2022simple} but has been recently extended to crystalline systems with the development of a novel functional~\cite{lu2024electron}.  Notably the combination of density functional theory with Macroscopic Quantum Electrodynamics has demonstrated the potential to treat realistic cavity setups accounting for excited state properties of materials in realistic lossy electromagnetic fields~\cite{svendsen2023molecules,Svendsen2021,scheel2009macroscopic,buhmann2012macroscopic,buhmann2013dispersion}. Finally, exact diagonalization of ab-initio based Hamiltonians represented in a reduced space has been used to predict the formation of composite quasi-particles of light~\cite{Latini2019, Latini2021} and hybrid light-matter groundstates~\cite{Latini2021-2}.

With the idea of cavity-material engineering gaining momentum, the ab initio theoretical treatment of a cavity coupled to an extended solid-state system needs to be formalized on common grounds, building on existing work~\cite{carusotto2013quantum, Vidal2021, bloch2022strongly, todorov2012intersubband}. Importantly for extended systems, it is crucial to account for the infinite amount of photonic degrees of freedom, inherent to the multi-mode nature of the electromagnetic field~\cite{amelio2021}, in order to avoid an artificial decoupling between light and matter when the bulk limit of the system is considered~\cite{lenk2022, Schlawin2019, mandal2023microscopic, taylor2024reciprocal}. Discarding such multi-mode nature would fail to capture light-matter interactions in bulk setups, contradicting recent experimental results~\cite{Thomas2019, jarc2023cavity}. The focus of this work is to demonstrate that it is possible to derive an effective ab initio theory of interacting light and matter in the groundstate based on an Hamiltonian consisting of few effective photonic degrees of freedom. Our method is hence distinct from existing effective approaches based on decoupled light-matter Green's function theory or which rely on external baths~\cite{Glauber1991dielectric, hughes2009effective, tamascelli2018nonperturbative}. The reason why an ab initio method is necessary, is to provide a framework where the validity of the theoretical predictions is not hindered by specific model choices, which has been the source of the different point of views in existing literature mentioned above~\cite{Vidal2021, bloch2022strongly}. More specifically, allowing for basis set convergence, an ab initio approach gives the opportunity to resolve potential issues with gauge ambiguities~\cite{de2018breakdown, di2019resolution, dmytruk2021gauge} arising from the fact that the matter system is described in a restricted basis set (such in the case of a tight-binding approximation). 
In doing so, we provide key conceptual advancements: (i) We show that even when neglecting the momentum carried by the light and considering an extended cavity the characteristic size of the confined effective electromagnetic radiation, and consequently the light-matter coupling, remains finite. (ii) We provide an Hamiltonian description grounded on QED free from the double counting of the coupling to the free-space electromagnetic background. (iii) We present the first fully Hamiltonian-based justification for the use of a few effective mode treatment of the cavity-matter problem at equilibrium. 

In practice, we show how the modes of a realistic Fabry-P\'erot cavity, i.e. with a non perfect reflectivity but loss-free, interact with an extended solid in the long-wavelength approximation employing the following steps: (a) We build the photonic modes as a linear combination of isotropic free-space basis states, solving Maxwell's equations for a finite reflectivity cavity (rather than introducing the cavity via idealized perfect boundary conditions). (b) We remove the spurious free-space light-matter coupling double counting by subtracting the contribution of the cavity in the limit of mirrors with zero reflectivity. (c) From the realistic cavity parameters and by employing the long wavelength approximation, we determine the relevant interaction length scales that define the strength of the light-matter coupling in the single effective mode treatment. Importantly, our scheme leads to an Hamiltonian which maintains the correct scaling properties with system size up to the limit of extended materials while featuring the simplicity of a few-photonic modes theory. Such an approach can be the starting point for numerically exact non-perturbative methods for simple systems ~\cite{sidler2022exact} as well as for more elaborate many-body approaches~\cite{schafer2022shining}, as discussed in the conclusion and outlook sections.

This work represents an important step towards the quantitative theoretical modeling of cavity-material engineering by connecting a realistic description of a cavity in the long wavelength approximation with the discretized momentum representation employed in modern electronic structure computational methods. 

\section{Light-Matter Coupling in QED}\label{sec:method_sec}
\subsection{Coupled light-matter Hamiltonian in free space}

The coupling between quantized matter and quantized light is formally described by the theory of quantum electrodynamics (QED). In order to allow for considerations on the scaling of light-matter interaction on system size~\cite{thirring2013quantum}, we keep the dependence on the latter explicit. Furthermore, we note that under the constraints of homogeneity and isotropy of space, the quantization volume for the electromagnetic field has to be chosen as a cube with edge length $L$ and periodic boundary conditions~\cite{ruggenthaler2022understanding}. It is important to emphasize that $L$ is a completely arbitrary length scale and therefore we should expect no physical dependence on its choice in the final theory.

In the Coulomb gauge, we can restrict the explicit quantization of the electromagnetic field to the two transverse polarizations. The mode functions are the free-space plane wave solutions to Maxwell's equations and the related transverse vector potential reads~\cite{greiner1996field} (we use SI-based atomic units throughout this work)
\be
\label{eq:vectorpotential}
    \hat{\bfA}_{\rm free}(\bfr) = \sqrt{\tfrac{2\pi}{ V}}\sum_{\bfq \lambda} \tfrac{\beps_{\bfq \lambda}}{\sqrt{ \omega_{\bfq}}} e^{\imagi \bfq \cdot \bfr }\left(\hat{a}_{\bfq \lambda} + \hat{a}_{-\bfq \lambda}^{\dagger}\right),
\ee
where $V=L^3$ is the quantization volume, $\omega_{\bfq}= c |\bfq|$ the frequency of the mode with momentum $\bfq$,  $\beps_{\bfq \lambda}$ the polarization function and $\lambda$ is an index that runs over the two transverse polarizations. The longitudinal part of the electromagnetic field leads instead to the well known matter-matter Coulomb interaction~\cite{greiner1996field}.

We introduce the light-matter coupling via the minimal-coupling prescription $\bf{\hat{p}} \rightarrow \bf{\hat{p}} + \bAh$ by imposing local gauge invariance. This ensures local charge conservation and in principle provides us with a fully relativistic description of QED. Here we use the low-energy approximation of QED as encoded in the Pauli-Fierz (PF) Hamiltonian~\cite{spohn2004dynamics},
\be\label{eq:paulifierz}
\begin{split}
\hat{H}_{\rm PF, free} =&  \sum_{l=1}^{N_e}\tfrac{1}{2}\left(-\imagi\bnabla_{l} \! + \! \hat{\bfA}_{\rm free}(\bfr_l) \right) ^2 +\frac{1}{2}\sum_{l \neq m}^{N_e} v(\bfr_l,\bfr_m) \\ 
&  +\sum_{l=1}^{N_e} \phi(\bfr_l) + \sum_{\bfq \lambda} \omega_{\bfq} \left(\hat{a}^{\dagger}_{\bfq\lambda}\hat{a}_{\bfq\lambda}+\frac{1}{2}\right),
\end{split}
\ee
%
where $v(\bfr_l,\bfr_m)$ and $\phi(\bfr_l)$ are the Coulomb interaction and the external potential acting on the electrons, respectively. Note that the ab initio PF Hamiltonian can be extended to include the nuclei/ions as effective quantum particles~\cite{Jestadt2019, ruggenthaler2022understanding} and to the (semi-)relativistic limit~\cite{spohn2004dynamics}. A key feature of the PF Hamiltonian is that it guarantees the existence of a groundstate and hence provides unambiguous access to equilibrium properties of coupled light-matter systems~\cite{schafer2020relevance}. It thus allows for extensions of various known first-principles methods from quantum mechanics to QED~\cite{ruggenthaler2022understanding}.

\subsection{Quantization of the electromagnetic field in the presence of mirrors: the Fabry-P\'erot cavity}

The relatively featureless electromagnetic modes of free space can be modified by introducing tailored electromagnetically active material structures in the environment. We will refer to such structures as a \textit{cavity}, the presence of which serves to alter the electromagnetic environment by selectively enhancing certain modes while suppressing others. In simple terms, the role of the cavity is to redistribute the electromagnetic states both spatially and spectrally. The resonances of a given cavity setup will depend on its material composition, size, and structural morphology~\cite{stockman2011nanoplasmonics,kuznetsov2016optically,mortensen2014generalized,svendsen2020role}.
There consequently exists a wide range of cavity designs, ranging from a simple planar geometry, to bow-tie shaped arrays, metasurfaces and photonic crystals with more complex topology and all this is in turn paired with the possibility of using both metallic and dielectric material constituents ~\cite{Miroshnichenko2022,Naylor2016,Deng2018,Abhiraman2020,Miao2020,kuznetsov2016optically,baranov2017all,svendsen2022computational}. Considering the vast degrees of freedom in cavity fabrication, both in the choice of the geometry and the constituting materials, and the fact that light-matter interaction is a joint light and matter property, the design space of cavity-matter systems is immense. \\

\begin{figure}[t]
    \centering
    \includegraphics[width=0.5\textwidth]{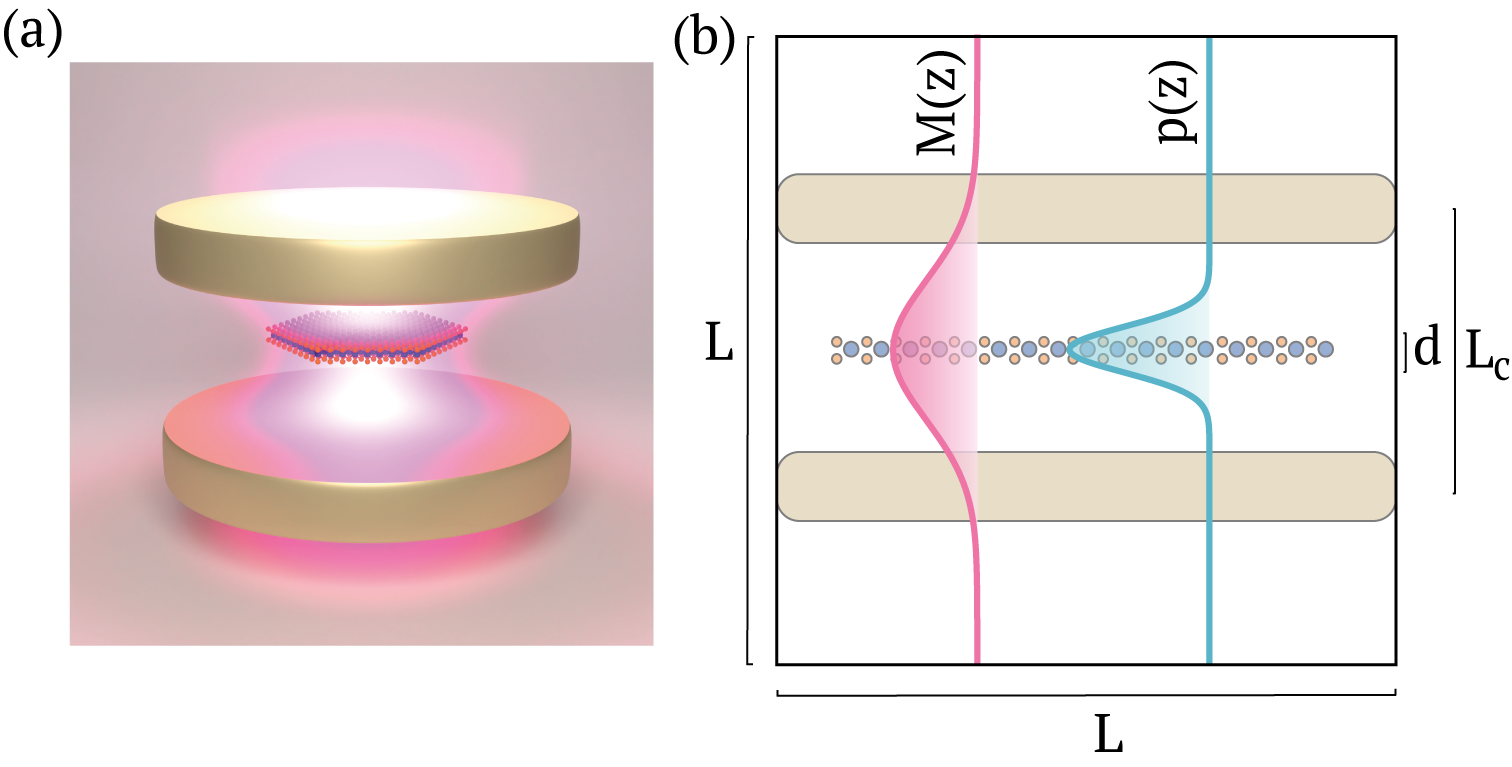}
    \caption{(a) Illustration of a realistic, non-perfectly reflecting Fabry-P\'erot cavity hosting a 2D crystal. (b) For the mathematical description, the cavity-matter system is contained in an isotropic, cubic quantization box with sides of length $L$. In out-of-plane ($z$)-direction the cavity and material have a fixed length scale, i.e., the mirrors are at a distance $L_{\rm c}$ and the material has a thickness $d$ and it is placed in the center of the cavity. To make the quantization procedure simple we choose a cavity, where the in-plane length scale coincides with the quantization length, i.e., $L_{\parallel} = L$. Panel (b) also shows a sketch of the out-of-plane dependence for the relevant fundamental mode function $M(z)$ of the photonic field and for the matter momentum matrix element $p(z)$.\label{fig:cavity_real}}
\end{figure}

In this work, we consider a paradigmatic example of an optical cavity consisting of two parallel thin mirrors separated by a distance $L_{\rm c}$ which can host a 2D extended material, as sketched in Fig.~\ref{fig:cavity_real}.
This setup is referred to as a Fabry-P\'erot cavity. The presence of the mirrors promotes the modes which are (near) compatible with the standing wave conditions, i.e. $q_z = n\frac{\pi}{L_{\rm c}},\, n\in\mathbb{N}$ and suppresses the rest. Such a setup is particularly amenable to asses the scaling of the light-matter coupling with system size as both the host material and the electromagnetic environment can be easily scaled in the planar dimensions. Unlike usually done in the literature, we introduce the cavity in the electromagnetic environment not by imposing perfect boundary conditions but rather by considering realistic mirrors whose reflectivity and transmission are characterized by the Fresnel coefficients, $r, t$. We assume that the cavity has no effect on the longitudinal part, i.e. that in the cavity the longitudinal Coulomb interaction is not affected (see Appendix~\ref{app:longitudinal} for a more detailed discussion). This approximation is reasonable for a material that is placed far enough from the cavity mirror, as the effect of longitudinal fields decay as a power law with the mirror distance~\cite{saez2023can}. We also specialize to a cavity with identical top and bottom mirrors while emphasizing that this choice has no qualitative impact on the results.
We can therefore directly expand the quantized vector potential in the cavity in terms of the original free-space photonic modes in Eq.~\eqref{eq:vectorpotential} provided that the mode functions are modified accordingly. 

In the following we fix the two thin mirrors at $z=\pm L_c/2$ respectively. The presence of the mirrors breaks translation invariance in the out-of-plane direction and couples modes propagating upwards and downwards. It is therefore convenient to separate the sum over the photonic modes in Eq.~\eqref{eq:vectorpotential} into in-plane and out-of-plane momenta sums and further split the latter into top propagating and bottom propagating momenta. A proper analytical expression for the cavity modified vector potential can be derived using standard transfer matrix techniques as shown in Appendix~\ref{app:fpc} and reads,
\be
\label{eq:AfieldFPC_all}
\begin{split}
    \bAh(\mathbf{r}) =& \frac{1}{\sqrt{V}}\sum_{\bfq_\parallel,\lambda}\sum_{q_z, \alpha} A_{0\bfq}e^{i\bfq_\parallel\cdot\mathbf{r}}(\bfM_{\bfq_\parallel\lambda, q_z\alpha}(z)\hat{a}_{\bfq_\parallel\lambda, q_z\alpha} + \\
    &\bfM_{-\bfq_\parallel\lambda, q_z\alpha}^*(z)\hat{a}^{\dagger}_{-\bfq_\parallel\lambda, q_z\alpha}).
\end{split}
\ee
The above expansion is rather intuitive, as the effect of the Fabry-P\'erot cavity is simply encoded in the $\bfM_{\bfq_\parallel\lambda, q_z\alpha}(z)$ mode functions, with $\lambda$ either the $s$ or $p$ polarization~\cite{saleh2019fundamentals} and $\alpha$ top and bottom propagation index. We emphasize that each mode is normalized over the full quantization volume $V=L^3$. Introducing the mirrors in the original quantization box breaks isotropicity, potentially challenging the possibility of recovering the isotropic free-space limit. We stress however that since we have control over the Fresnel coefficients of the mirror we can recover the correct free space limit by making the mirrors completely transparent. It is important to mention that the quantization procedure here is equivalent to existing procedure used in the context of Macroscopic QED in the absence of losses~\cite{Glauber1991dielectric}. \\

\subsection{Coupled light-matter Hamiltonian with a Fabry P\'erot cavity}

Next we consider, as a realistic, lossless, prototypical cavity-matter system, a Fabry-P\'erot cavity hosting a 2D crystal, as sketched in Fig.~\ref{fig:cavity_real}. The discussion can be readily adapted to any type of lossless cavity. For notational convenience, we first translate the PF Hamiltonian into the language of second quantization for the electrons. It is important to stress that for any calculation employing the PF Hamiltonian in second quantization, without further care, one should restrict to a specific particle-number subspace of the full Fock space to have physically meaningful results (see Appendix~\ref{app:grandcanonical} for further details).
For extended 2D crystalline systems, the in-plane periodicity of the crystal potential $\phi(\bfr) = \phi(\bfr + \bfR_\parallel)$ makes it convenient to represent the fermionic annihilation and creation operators in terms of in-plane periodic Bloch's functions,%
\begin{align}
\hat{\psi}(\bfr \sigma) = \tfrac{1}{\sqrt{S_{\rm M}}}\sum_{i\bfk_\parallel}^{\rm BZ}e^{\imagi\bfk_\parallel\cdot(\tilde{\bfr} + \bfR_\parallel) }u_{i\bfk_\parallel}(\tilde{\bfr} \sigma)\hat{c}_{i\bfk_\parallel\sigma},
\end{align}
with $\tilde{\bfr}$ restricted to the unit cell, $\bfR_\parallel$ the lattice vector of the $n$-th unit cell, $u_{i\bfk_\parallel}(\tilde{\bfr} \sigma) = u_{i\bfk_\parallel}((\tilde{\bfr}+ \bfR_\parallel) \sigma)$ is the periodic part of the Bloch wave function, $S_{\rm M}$ the in-plane area covered by the full matter system, and $\hat{c}_{i\bfk_\parallel \sigma}$ the annihilation operator of an electron in a Bloch state with index $i\bfk_\parallel$ in the first Brillouin zone (BZ) of the crystal. In general, the periodicity of the crystal might not be consistent with the cubic symmetry implied by the quantized electromagnetic field. In such cases there is no direct mapping between matter and photon momenta. A discussion on this and related issues is presented in Appendix~\ref{app:mismatch}. Crucially, the \ac{LWA} will allow us to avoid this problem.
Under the assumptions mentioned above, and as shown in Appendix~\ref{app:PFBloch} and similar to Ref.~\cite{amelio2021}, the Bloch form of the PF Hamiltonian for the coupled Fabry-P\'erot cavity matter system becomes,
\begin{widetext}
\be
\label{eq:multimodes}
\begin{split}
\hat{H}_\mathrm{PF} =& \hat{H}_\mathrm{EM} + \hat{H}_\mathrm{el} + \frac{1}{\sqrt{V}}\sum_{ij\sigma\bfk_{\parallel}}\sum_{\bfq_\parallel,\lambda}\sum_{q_z, \alpha}\int_{\Omega_{z}}dz~ \hat{c}_{i\bfk_\parallel+ \bfq_\parallel\sigma}^\dagger\hat{c}_{j\bfk_\parallel\sigma} A_{0\bfq_\parallel q_z}\bfp_{ij\sigma\bfk_{\parallel}\bfq_{\parallel} q_z}(z)\cdot[\bfM_{\bfq_\parallel \lambda, q_z \alpha}(z)\hat{a}_{\bfq_\parallel\lambda,q_z\alpha} \\
&+\bfM_{-\bfq_\parallel\lambda,q_z\alpha}^*(z)\hat{a}^{\dagger}_{-\bfq_\parallel\lambda,q_z\alpha}]+ \frac{1}{2V}\sum_{ij\sigma\bfk_\parallel}\sum_{\bfq_\parallel\bfq_\parallel',\lambda\lambda'}\sum_{q_z q_z',\alpha\alpha'}\int_{\Omega_{z}}dz~\hat{c}_{i\bfk_\parallel+\bfq_\parallel+\bfq_\parallel'\sigma}^\dagger\hat{c}_{j\bfk_\parallel\sigma} A_{0\bfq_\parallel q_z}A_{0\bfq_\parallel'q_z'}s_{ij\sigma\bfk_\parallel\bfq_\parallel\bfq_\parallel'}(z)\\
&[\bfM_{\bfq_\parallel\lambda,q_z\alpha}(z)\hat{a}_{\bfq_\parallel\lambda q_z\alpha} +\bfM_{-\bfq_\parallel\lambda,q_z\alpha}^*(z)\hat{a}^{\dagger}_{-\bfq_\parallel\lambda,q_z\alpha}]\cdot[\bfM_{\bfq_\parallel'\lambda',q_z'\alpha'}(z)\hat{a}_{\bfq_\parallel'\lambda',q_z'\alpha'} +\bfM_{-\bfq_\parallel'\lambda',q_z'\alpha'}^*(z)\hat{a}^{\dagger}_{-\bfq_\parallel'\lambda',q_z'\alpha'}],
\end{split}
\ee
\end{widetext}
where we have grouped the electronic single-particle and interaction terms into $\hat{H}_{\rm el}$, defined $\Omega_{z}$ as the matter unit cell dimension in the z-direction (note that since here we consider no momentum dispersion for the matter in the z-direction, the unit cell should contain all the potential layers of the material) and we have defined the momentum and overlap matrix elements, $\bfp_{ij\sigma\bfk_{\parallel}\bfq_{\parallel} q_z}(z)$ and $s_{ij\sigma\bfk_\parallel\bfq_\parallel\bfq_\parallel'}(z)$ respectively, in Appendix~\ref{app:PFBloch}. Eq.~\eqref{eq:multimodes} explicitly contains all the possible momentum conserving electron-photon interactions separated in a paramagnetic (coupling to the matter momenta) and a diamagnetic (coupling to the matter density) term. It is interesting to notice how the diamagnetic term written in this general multi-mode framework explicitly couples otherwise non-interacting photonic modes because of the presence of matter.

\section{The long wavelength approximation for the coupled light-matter system}\label{sec:LWA}

In its full form, the PF Hamiltonian contains a potentially infinite amount of cavity modes, and these modes are mutually coupled via the diamagnetic term. This makes working with the full PF Hamiltonian highly impractical. 
In the following we will therefore seek a suitable simplification which will allow us to consider the effect of the cavity on a material in its bulk limit by applying the long wavelength approximation.

\subsection{Operational Definitions}
Because the field of interacting light-matter systems is approached by experts with diverse backgrounds, e.g. condensed matter physics, quantum optics and quantum chemistry, we find it essential to provide a working definition of the nomenclature used throughout this work.  
Whenever referring to the long wavelength approximation~(LWA), we assume that the momentum carried by the modes of the electromagnetic field can be neglected from the point of view of the matter. The validity of this assumption is guaranteed as long as the photon momenta are below some momentum cut-offs for the in-plane-, and out-of-plane components, denoted by $\bfq_{\rm c,\parallel}^{\rm lw}$ and $q_{\rm c, z}^{\rm lw}$, respectively. 
The LWA has much in common with the widely used \textit{dipole approximation} of atomic and molecular physics. However, while they are the same for finite systems like atoms and molecules, the two have important differences when considering extended systems. Within the dipole approximation, it is assumed that the spatial extent of the matter is much smaller than the characteristic length scale of the field variations. Any mode with $|\bfq_\parallel| > \frac{2\pi}{S_M^{1/2}}$ should therefore be discarded. As we increase the size of the system we thus have to discard more and more modes from our theory to fulfil the dipole approximation, effectively leading to a light-matter decoupling in the limit of a bulk material. Within this definition of dipole approximation a number of recent works~\cite{Eckhardt2022, Andolina2019}, have argued that in the limit where the cavity-matter size goes to infinity, the effect of the  virtual fluctuations of the cavity electromagnetic field on the material properties vanishes. However we stress that this decoupling reflects the stringent conditions enforced by the dipole approximation which ignores the multi-mode nature of the electromagnetic field. 
Especially in the case of extended systems, the LWA is instead less stringent~\cite{amelio2021,lenk2022}. Disregarding surface modes such as surface plasmon polaritons, which are beyond the applicability of the approach in this work, the momentum carried by any mode in a Fabry-P\'erot cavity is negligible compared to the momentum scales over which the electronic matrix elements vary. For the momentum transfers that actually enter into the sums in Eq.~\ref{eq:multimodes}, we are justified in applying the LWA even though $|\bfq_\parallel| > \frac{2\pi}{S_M^{1/2}}$ for some of the modes up to the cut-offs defined above. \\

In view of the analysis on the scaling of the light-matter coupling with system size it is relevant to discuss the definition of the \textit{thermodynamic limit}. Given a system of volume $V$, the thermodynamic limit (often referred to as \textit{macroscopic limit}) is defined as $V\rightarrow \infty$. While taking this limit is formally justified when treating the uncoupled light or the uncoupled matter systems alone, the same is not automatically true for hybrid-light matter systems as discussed in more details in Sec.~\ref{sec:macroscopic_limit_and_length_scales_of_qed}. In short, for the strict dipole approximation if we take the volume to infinity we decouple light and matter; on the contrary, within the LWA, if the amount of matter included in the coupling is not restricted (by the realistic length scales set by the cavity as shown later in this work) the light matter coupling diverges to infinity. Therefore, in this work we consider instead the \textit{bulk limit} which is defined as the limit where the material properties are the ones of its bulk form, yet below the characteristic lengths set by the cavity as explained in Sec.~\ref{sec:macroscopic_limit_and_length_scales_of_qed}. This definition is introduced for ab initio calculations in Sec.Sec.~\ref{sec:convergence}. Note that since in this work we consider 2D crystals, the bulk limit should be interpreted as the limit of a 2D bulk slab. 

\subsection{PF Hamiltonian in the LWA}

To make the effective \ac{LWA} of Eq.~\eqref{eq:multimodes}, we can neglect the variation of the electromagnetic field across the vertical width $d$ of the 2D material and define the $\bfq_\parallel = 0$ component of $\bar{\bfp}_{ij\sigma\bfk\bfq}(z)$ integrated in the z-direction as $\bar{\bfp}_{ij\sigma\bfk\boldsymbol{0}}$ (see Appendix~\ref{app:PFBloch}). We note that for thicker materials with widths comparable to the wavelength of the cavity mode, it will be important to consider the variation of the field across the width of the material. However, that will not qualitatively impact the conclusions in the following. This allows us to evaluate the polarization functions of the cavity directly at its center and it is therefore convenient to define $\bar{\bfM}_{\bfq_\parallel\lambda,q_z\alpha}=\bfM_{\bfq_\parallel\lambda,q_z\alpha}(z=0)$. We can then rewrite the photon mode functions as
    $\bar{\bfM}_{\bfq_\parallel\lambda,q_z\alpha} = \bar{\bfM}_{\bfq_\parallel\lambda,q_z\alpha}^\parallel + \bar{\bfM}_{\bfq_\parallel\lambda,q_z\alpha}^z$,
where $\parallel$($z$) refers to the part of $\bar{\bfM}_{\bfq_\parallel\lambda,q_z\alpha}$ which is parallel(perpendicular) to the mirrors. Under the approximation of vanishing out-of-plane polarization of the matter, we can thus write,
\begin{align}
    \bar{\bfp}_{ij\sigma\bfk\boldsymbol{0}}\cdot \bar{\bfM}_{\bfq_\parallel,s,q_z\alpha} = \bar{\bfp}_{ij\sigma\bfk\boldsymbol{0}}\cdot \bar{\bfM}_{\bfq_\parallel,s,q_z\alpha}^\parallel.
\end{align}
By inspection of the polarization vectors in Eqs.~\eqref{app:eq:polarization:s}-\eqref{app:eq:polarization:pb} we see that the $s$-polarization is unaffected by the approximation since $\bar{\bfM}_{\bfq_\parallel,s,q_z\alpha} = \bar{\bfM}_{\bfq_\parallel,s,q_z\alpha}^\parallel$ already holds, unlike the case of the $p$-polarization. As shown in Appendix~\ref{app:fpc}, for both polarizations we can define a new set of photonic operators,
\begin{align}\label{eq:new_b_ops}
    &
    \hat{b}_{\bfq_\parallel \lambda q_z\alpha} \equiv \frac{\bar{R}_{\bfq_\parallel,\lambda,q_z\alpha}}{|\bar{R}_{\bfq_\parallel,\lambda,q_z\alpha}|} \hat{a}_{\bfq_\parallel \lambda q_z\alpha}, \\
    &\hat{b}^\dagger_{\bfq_\parallel \lambda q_z\alpha} \equiv \frac{\bar{R}_{\bfq_\parallel,\lambda,q_z\alpha}}{|\bar{R}_{\bfq_\parallel,\lambda,q_z\alpha}|} \hat{a}_{\bfq_\parallel \lambda q_z\alpha}^\dagger\label{eq:new_b_ops_dagger},
\end{align}
and new effective vector-potential amplitudes for the two polarizations,
\begin{align}\label{eq:new_A_amps_s}
    &\bar{\bfA}_{\bfq_\parallel s, q_z\alpha} = A_{0,\bfq_\parallel q_z}\left|\bar{R}_{\bfq_\parallel,s,q_z\alpha}\right|\beps^\parallel_{\bfq_\parallel,s, q_z},\\
    &\bar{\bfA}_{\bfq_\parallel p, q_z\alpha} = \frac{q_z}{\sqrt{q_z^2 + \bfq_\parallel^2}}A_{0,\bfq_\parallel q_z}\left|\bar{R}_{\bfq_\parallel,p,q_z\alpha}\right|\beps^\parallel_{\bfq_\parallel,p, q_z}\label{eq:new_A_amps_p},
\end{align}
where $\bar{R}_{\bfq_\parallel,p,q_z\alpha}$ are the new mode functions as derived in Appendix~\ref{app:fpc}. Notice that the prefactor $\frac{q_z}{\sqrt{q_z^2 + \bfq_\parallel^2}}$ takes care of recovering the purely photonic effect of the out-of-plane component of the $p$-polarization mode function which would otherwise be lost (see App.~\ref{app:splittingpol}). If we assume that the polarization of the electronic system is isotropic in-plane, the product $\bar{\bfp}_{ij\sigma\bfk\boldsymbol{0}}\cdot\beps_{\bfq_\parallel,\lambda, q_z}$ is the same for all directions. We can therefore define $\bar{\bfp}_{ij\sigma\bfk\boldsymbol{0}}\cdot\beps_{\bfq_\parallel,\lambda, q_z} \equiv \bar{p}_{ij\sigma\bfk\boldsymbol{0}}$ and recast Eq.~\eqref{eq:multimodes} as,
\begin{widetext}
\be
\label{eq:hamlw}
\begin{split}
    \hat{H}^\mathrm{lw}_\mathrm{PF} &\simeq \hat{H}_\mathrm{EM} + \hat{H}_\mathrm{el}
    + \frac{1}{\sqrt{V}}\sum_{ij\sigma\bfk_\parallel}\hat{c}_{i\bfk_\parallel\sigma}^\dagger\hat{c}_{j\bfk_\parallel\sigma}\bar{p}_{ij\sigma\bfk\boldsymbol{0}}\sum_{\bfq_\parallel,\lambda}^{\bfq_{\rm c,\parallel}^{\rm lw}}\sum_{q_z,\alpha}^{\bfq_{\rm c, z}^{\rm lw}} \bar{A}_{\bfq_\parallel\lambda, q_z\alpha}(\hat{b}_{\bfq_\parallel\lambda,q_z\alpha}+\hat{b}^{\dagger}_{-\bfq_\parallel\lambda,q_z\alpha})\\
    &+ \frac{1}{2V}\sum_{i\sigma\bfk_\parallel}\hat{c}_{i\bfk_\parallel\sigma}^\dagger\hat{c}_{i\bfk_\parallel\sigma}\sum_{\bfq_\parallel,\lambda}^{\bfq_{\rm c,\parallel}^{\rm lw}}\sum_{q_z,\alpha}^{\bfq_{\rm c,z}^{\rm lw}}\bar{A}_{\bfq_\parallel\lambda, q_z\alpha}^2(\hat{b}_{\bfq_\parallel\lambda,q_z\alpha} +\hat{b}^{\dagger}_{-\bfq_\parallel\lambda,q_z\alpha})^2.
\end{split}
\ee
\end{widetext}

\begin{figure*}[t]
    \centering
    \includegraphics[width=0.95\textwidth]{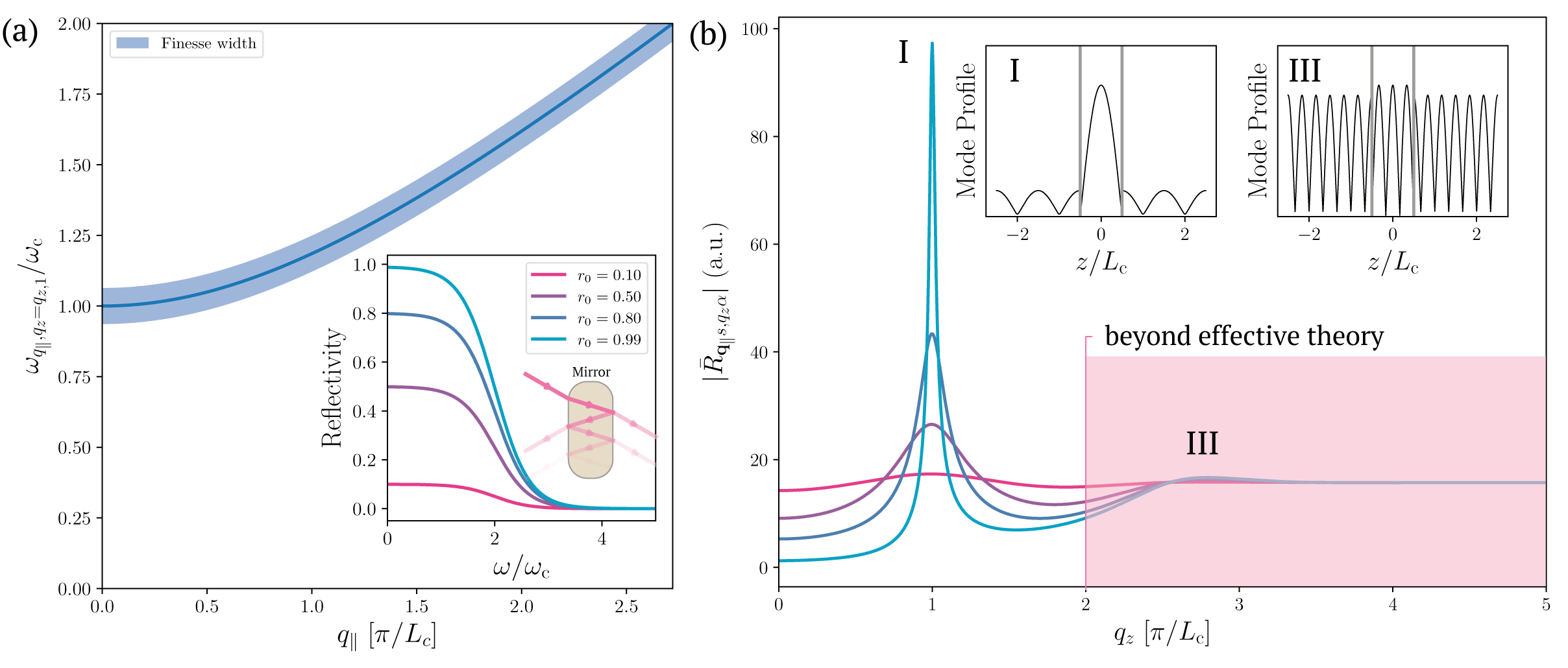}
    \caption{(a) Mode dispersion of the fundamental mode in the ideal Fabry-P\'erot cavity, and (b) the mode function of the ideal Fabry-P\'erot cavity in $q_z$-space.
    The effect of the cavity is to pin $q_z$ to a narrow range, and the result is that the relevant frequencies to count in the effective mode construction are the ones shown by the shaded area in (a). The width of this area depends on the width $\gamma$ which in turn is related to the mirror reflectivity as discussed in Sec.~\ref{subsec:out_of_plane_cutoff}. The inset in (a) shows different reflectivity profiles used to generate the curves in (b). Modes above the reflectivity cut-off quickly approach the free-space case. This is also clear when looking at the spatial mode functions in the inset of (b) which shows the first and third mode for a reflectivity $r = 0.95$. Since the first mode lives in the frequency region where the mirrors are highly reflective and the the third lives above the plasma frequency of the mirror, $\omega_p$, the confinement of the first mode is significantly larger than for the third mode and it is consequently closer to a standing wave and significantly more enhanced.}
    \label{fig:cavity_functions}
\end{figure*}

To obtain the equations above we have also used the fact that the matrix elements $s_{ij\sigma \bfk_\parallel \boldsymbol{00}}= \delta_{ij}$ due to the orthonormality of the $u_{i\bfk_\parallel}(\tilde{\bfr}\sigma)$ on the unit cell. We therefore observe that the effect of the cavity is to ``dress" the free-space modes of the electromagnetic field by changing the prominence of different wave vectors in the cavity. Specifically, as shown in Fig.~\ref{fig:cavity_functions}(b), from the dependence of $\left|\bar{R}_{\bfq_\parallel,p,q_z\alpha}\right|$ on $q_z$, we observe that the cavity enhances $q_z$ which are close to the standing wave condition in the limit of the perfectly reflective Fabry-P\'erot cavity $q_{c,n} = n \frac{\pi}{L_c}$ for $n\in\mathbb{N}$, and suppress the rest. We only observe the odd modes in Fig.\ref{fig:cavity_functions}(b) because the material is placed in the center of the cavity and thus overlaps with a node of the even modes and an anti-node of the odd modes. This is consistent with what we expect from the real- and reciprocal space characteristics of a Fabry-P\'erot cavity.

A remaining question is what cut-offs to apply for the in-plane and out-of-plane directions. We will return to this point in sections~\ref{subsubsec:inplanecutoff} and~\ref{subsec:out_of_plane_cutoff} respectively.

\subsection{Removing free-space contributions from the light-matter coupling}
\label{subsubsec:removefree}

A final important issue for any calculation involving light-matter coupling from QED is the comparison to the free-space limit. Physically one needs to make sure that coupling of matter to the fluctuating photons present in free space, which is already encoded in the observable masses of the particles, are not double counted in the theory. For this purpose our formulation of the light-matter problem is very convenient, indeed (i) it quantizes the electromagnetic field isotropically, which is the underlying assumption used in QED to derive the mass of the particles, (ii) it allows to recover the isotropic free-space limit of quantum mechanics by setting the reflectivity of the cavity to zero. Without such construction one would have to rewrite the PF Hamiltonian using bare particle masses which can become cumbersome. Our approach, instead, allows us to work with standard effective masses for the matter particles provided that we subtract the terms that are already present in the free-space coupling.

In order to accomplish this, we need to distinguish between the vacuum and reflected contributions to the photonic field. Looking at Eqs.~\eqref{eq:new_b_ops}-\eqref{eq:new_A_amps_p}, we observe that the effect of the cavity enters via $\bar{R}_{\bfq_\parallel,p,q_z\alpha}$. Noting that $\frac{1}{1-x}=\sum_{n=0}^\infty x^n$ for $|x|\leq 1$ we can write,
\begin{align}
    &\bar{R}_{\bfq_\parallel,\lambda,q_z\alpha} =\nonumber\\ &t\left[1+\left(re^{iq_zL_c}\sum_{n=0}^\infty(r^2e^{2iq_zL_c})^n + \sum_{n=1}^\infty(r^2e^{2iq_zL_c})^n\right)\right]
\end{align}
We approach free space by making the mirrors more transparent ($t\rightarrow 1$ and $r \rightarrow 0$). The first term contains no reflections and thus represents the remaining free space contribution while the last two terms are exclusively cavity contributions. In free space $\bar{R}_{\bfq_\parallel,\lambda,q_z\alpha} = 1$ and we can thus remove the free space double counting by using,
\begin{align}
    &\bar{R}^\mathrm{NF}_{\bfq_\parallel,\lambda,q_z\alpha} =\nonumber\\ &(t-1) + t\left[\left(re^{iq_zL_c}\sum_{n=0}^\infty(r^2e^{2iq_zL_c})^n + \sum_{n=1}^\infty(r^2e^{2iq_zL_c})^n\right)\right]
\end{align}
instead of the bare mode functions. In this way, we arrive at an effective mode construction which vanishes as we approach free space ($t\rightarrow 1$ and $r \rightarrow 0$) and the sole effect is to modify the effective mode function as in Fig.~\ref{fig:eff_coupling_strenths}(c). In the following, our definitions will therefore be in terms of $\bar{R}^\mathrm{NF}_{\bfq_\parallel,\lambda,q_z\alpha}$ and not $\bar{R}_{\bfq_\parallel,\lambda,q_z\alpha}$.

\section{The effective-mode description}

While significantly simpler than Eq.~\eqref{eq:multimodes}, the multi-mode nature of Eq.~\eqref{eq:hamlw}, which for the arbitrary length scale $L \rightarrow \infty$ would turn into a continuum, is still impractical for most calculations. In this section we devise an effective-mode description which keeps the appropriate size scaling of the light-matter problem while simplifying computations. The overarching idea is that within the \ac{LWA}, the effect of the photonic modes can be encoded into a single re-scaled photonic mode for each polarization consisting of collectively oscillating photons. Here we will highlight the main points leading to the effective-mode Hamiltonian and we refer to Appendix~\ref{app:singleeffective} for a step-by-step derivation. 

\subsection{Collective Canonical Transformation}\label{sec:canonical_transformation}

We first notice that if we neglect the angular ($\bfq_\parallel$) dependence of the mirror's Fresnel coefficients, $\bar{R}_{\bfq_\parallel,\lambda,q_z\alpha}$ is independent of $\bfq_\parallel$. For cavity mirrors with high reflectivity and a material placed in the center of the cavity, the functions $\bar{R}_{\bfq_\parallel,q_z\alpha}$ are peaked around the odd standing wave conditions of the perfectly reflecting Fabry-P\'erot cavity. Furthermore, for a real mirror, the frequency dependence of the reflectivity can be engineered to give a decreased confinement for modes with $n>1$ (see inset of Fig.~\ref{fig:cavity_functions}(a)) so that $\bar{A}_{\boldsymbol{0},q_z\alpha}$ can be approximated to a single peak function centered around $q_{z, 1}$ (like in Fig.~\ref{fig:cavity_functions}(b)). We stress that the latter is a convenience choice and that it would be possible to perform a similar procedure, separately for each of the modes (allowed by the cavity and the LWA) with $n>1$ and eventually arriving at an effective few modes Hamiltonian. In Fig.~\ref{fig:cavity_functions}(b) the out-of-plane mode profile associated with the visible peaks at $q_{z,1}$ and $q_{z,3}$ are illustrated in the insets to highlight how the non-perfect reflectivity of the mirrors allows ``leakage" outside the cavity. The "leakage" is exactly what allows to couple a cavity with radiation from the ouside. We note in passing that for any Fabry-P\'erot cavity, $q_{z, 1}<q_{{\rm c}, z}^{\rm lw}$ due to the atomic thickness of 2D materials guaranteeing consistency with the \ac{LWA} for the modes included in our theory. 

While $\bar{R}_{\bfq_\parallel,\lambda,q_z\alpha}$ does not depend on the in-plane momentum, the functions in Eqs.~\eqref{eq:new_A_amps_s} and ~\eqref{eq:new_A_amps_p}  do, via $\omega_{q_\parallel q_z} = \sqrt{q_\parallel^2+q_z^2}$. The behaviour of $\omega_{q_\parallel,q_z}$ when $q_z$ is pinned to the first cavity resonance is shown in Fig.~\ref{fig:cavity_functions}(a). We note that this dispersion, while critical when considering resonant phenomena, such as in the case of polariton relaxation~\cite{tichauer2021multi} and transport~\cite{berghuis2022controlling, xu2023ultrafast, balasubrahmaniyam2023enhanced}, is not expected to play an important role when considering equilibrium changes to the material due to the off-resonant nature of the coupling between light and matter in this case.

With all the above consideration we define an effective coefficient $A_{{\rm eff}, \lambda}$ by setting $\omega_{q_\parallel q_z}=\omega_{\rm c}$ and average the functions $\bar{A}_{\bfq_\parallel \lambda, q_z\alpha}$ in the $q_z$ variable within a region of width $\gamma$ centered around $q_{z,1}$, in formulas
\be\label{eq:effective_A}
A_{\rm eff, \lambda}\equiv \frac{1}{l_z}\sum_{q_z = q_{z,1}-\gamma/2}^{q_z = q_{z,1}+\gamma/2}\bar{A}_{q_\parallel=\sqrt{(\omega_{\rm c}/c)^2 - q_z^2} \lambda,q_z\alpha},
\ee
where we further note that $\bar{A}_{q_\parallel \lambda,q_z\alpha}$ is independent of the propagation direction $\alpha$ because the crystal is centered at $z = 0$. Note that $l_z$ here is the number of modes within the averaging region and together with $\gamma$ are formally defined in relation to the cut-offs as given in Sec.~\ref{subsec:out_of_plane_cutoff}.

Now that all modes share the same coefficient $A_{\rm eff, \lambda}$, it is natural to define the total displacement operator for each of the two polarizations,
\be
\label{eq:totdispl}
\hat{Q}_{{\rm eff}, \lambda} \equiv \frac{1}{\sqrt{2}} \sum_{\alpha}\sum_{\bfq_\parallel}^{\bfq_{{\rm c}, \parallel}^{\rm lw}}\sum_{q_z}^{\bfq_{{\rm c},z}^{\rm lw}} (\hat{b}_{\bfq_\parallel\lambda,q_z\alpha}+\hat{b}^{\dagger}_{-\bfq_\parallel\lambda,q_z\alpha}),
\ee
which is the photonic operator that directly appears in the paramagnetic term in Eq.~\eqref{eq:hamlw}. As shown in Appendix~\ref{app:singleeffective}, this total operator paired with the respective relative operators can be used as a starting point for a canonical transformation and allow us to simplify the \ac{LWA} PF Hamiltonian to,
\begin{widetext}
\begin{align}
\label{eq:lwsinglefinal}
\hat{H}_{\rm PF}^{\rm lw} \simeq& \hat{H}_{\rm el}+\hat{H}_{\rm EM, rel}+\sum_\lambda\left(\omega_{\rm c, eff}+\frac{1}{2}\right)\hat{B}^{\dagger}_{\rm eff, \lambda}\hat{B}_{\rm eff, \lambda} + \nonumber \\ &\left(\frac{l_\parallel l_z}{V}\right)^{1/2}\hat{P}_{\rm el} \sum_\lambda A_{\rm eff,\lambda} \left(\hat{B}^{\dagger}_{\rm eff, \lambda}+\hat{B}_{\rm eff, \lambda}\right) + \left(\frac{l_\parallel l_z}{V}\right)\hat{N}_{\rm el}\frac{A_{\rm eff, \lambda}^2}{2}\left(\hat{B}^{\dagger}_{\rm eff, \lambda}+\hat{B}_{\rm eff, \lambda}\right)^2,
\end{align}
\end{widetext}
where we have introduced a new set of creation and annihilation photonic operators $\hat{B}^{\dagger}_{\rm eff}$ and $\hat{B}_{\rm eff}$ associated with the operator $\hat{Q}_{{\rm eff}, \lambda}$, defined the total electron momentum operator as $\hat{P}_{\rm el}\equiv\sum_{ij\sigma\bfk_\parallel}\hat{c}_{i\bfk_\parallel\sigma}^\dagger\hat{c}_{j\bfk_\parallel\sigma}\bar{p}_{ij\sigma\bfk\boldsymbol{0}}$, the electronic number operator, and $l_{\parallel}$ is defined in Sec.~\ref{subsubsec:inplanecutoff}. We note that the total momentum and number of particles operators are not bound to the effective-mode construction but they already appear in Eq.~\eqref{eq:hamlw} due to the \ac{LWA}. Note further that we have grouped the terms related to the relative photonic coordinates into $\hat{H}_{\rm EM, rel}$.

Eq.~\eqref{eq:lwsinglefinal} directly highlights a fundamental feature of our effective theory: the collective photon modes acquire the weight of all the original photonic modes up to the cut-offs $\bfq_{\rm c, \parallel}^{\rm lw}$ and  $\bfq_{\rm c, z}^{\rm lw}$. To determine how the light-matter coupling scales with quantization volume it is therefore necessary to determine how many modes contribute to the effective mode construction. As we shall see, this indeed leads to a light-matter coupling which is \textit{independent} of the quantization volume.

\subsection{In-Plane cut-off}
\label{subsubsec:inplanecutoff}
If we neglect coupling to any surface modes that might exist near the mirrors, the largest momentum of any mode modified by the Fabry-P\'erot cavity is $q_p = \frac{\omega_{\rm p}}{c}$ where $\omega_{\rm p}$ is the frequency where the mirrors reflectivity drops (which in the case of metallic mirrors would roughly be the plasma frequency). Within the strict dipole approximation, we therefore wrongly discard all modes with $2\pi/L_M\leq q_\parallel \leq q_p$ from the light-matter coupling. Discarding these modes means that for an extended system the coupling to the majority of the modes in the cavity is not described and one there cannot expect the theory to yield physically correct results.

To formally define the cut-off in the in-plane direction, we note that while the cavity modifies the prominence of different $q_z$, it has no direct impact on $q_\parallel$. We can therefore uniformly sum the contribution of all the $q_\parallel$ in the cavity such that $q_\parallel \leq q_p$. Because fixing the frequency means that $q_\parallel = \sqrt{\left(\frac{\omega}{c}\right)^2 - q_z^2}$ and the cavity pins $q_z$ to $q_{z,1}$, we can define the in-plane momentum cut-off as $q_{c,\parallel}^{\rm lw} = \sqrt{\left(\frac{\omega_p}{c}\right)^2 - q_{z,1}^2}$. Note that for our cavity to only host a single mode we should choose $\omega_p = 2\omega_1$ where $\omega_1 = \frac{q_{z,1}}{c}$. This way, it appears that all modes with $|q_\parallel|\in[0,\sqrt{3}q_{z,1}]$ have to be counted in the coupling. Taking into account that the spacing between photon modes is set by the quantization-box size $L$, we get that the number of photon modes $l_\parallel$ is given by
\be
\label{eq:pcut}
l_{\parallel} = \left(\frac{q^\mathrm{lw}_{c,\parallel}}{\Delta q}\right)^2 = \frac{\sqrt{3}q_{z,1}^2A}{4\pi^2},
\ee
where $\Delta q = (2\pi)^2/A^2$ is the area of each momentum grid element.
Inserting into Eq.~\ref{eq:lwsinglefinal}, we see that the quantization-box size $L$ does not change the strength of the in-plane cavity coupling. Instead, by increasing $L$ we merely find more and more modes carrying less and less individual weight such that we have a well-defined continuum limit. The same holds true also for the out-of-plane modes. \\

\subsection{The Out-of-Plane cut-off}\label{subsec:out_of_plane_cutoff}

As discussed in Sec.\ref{sec:canonical_transformation} the sum over modes in the out-of-plane direction is set by the width of the mode function with respect to $q_z$, i.e. $\gamma$. In the Fabry-P\'erot cavity description used in this work, this can be calculated once the reflectivity of the mirror is defined. This is thus a cavity specific quantity. Specifically, for relatively large mirror reflectivity where the individual modes of the Fabry-P\'erot cavity can be resolved, we can meaningfully approximate $\gamma$ as~\cite{saleh2019fundamentals},
\begin{align}
    \gamma \approx \frac{1}{\mathcal{F}}\frac{\pi}{L_c}
\end{align}
where the cavity finesse is $\mathcal{F} = -\frac{2\pi}{\mathrm{ln}(|r|^4)}$. The value of $\gamma$ is in turn related to the number of modes $l_{z}$ by
\be
\label{eq:zcut}
l_{z} = \frac{\gamma}{\Delta q} = \frac{\gamma L}{2\pi} \approx \frac{1}{\mathcal{F}}\frac{L}{2L_c}.
\ee
With the last relation we have fully connected the light-matter coupling parameters in the effective-mode theory to the specific characteristics of the cavity device. Together with the in-plane coupling we find that in Eq.~\eqref{eq:lwsinglefinal} the coupling becomes independent of the quantization volume $V=L^3$. We note that Eq.~\eqref{eq:zcut} suggests that the light-matter coupling formally vanishes when $r\rightarrow 1$ and in turn $\mathcal{F}\rightarrow\infty$. However, at $r\rightarrow 1$, $A_{\mathrm{eff},\lambda}$ is also divergent and so one needs to take this limit carefully. 

The results obtained above are specific to the Fabry-P\'erot cavity, however they highlight the importance of light and matter length scales and mirror characteristics when defining the problem for the coupled system. 
\begin{figure*}[t!]
    \centering
    \includegraphics[width=1\textwidth]{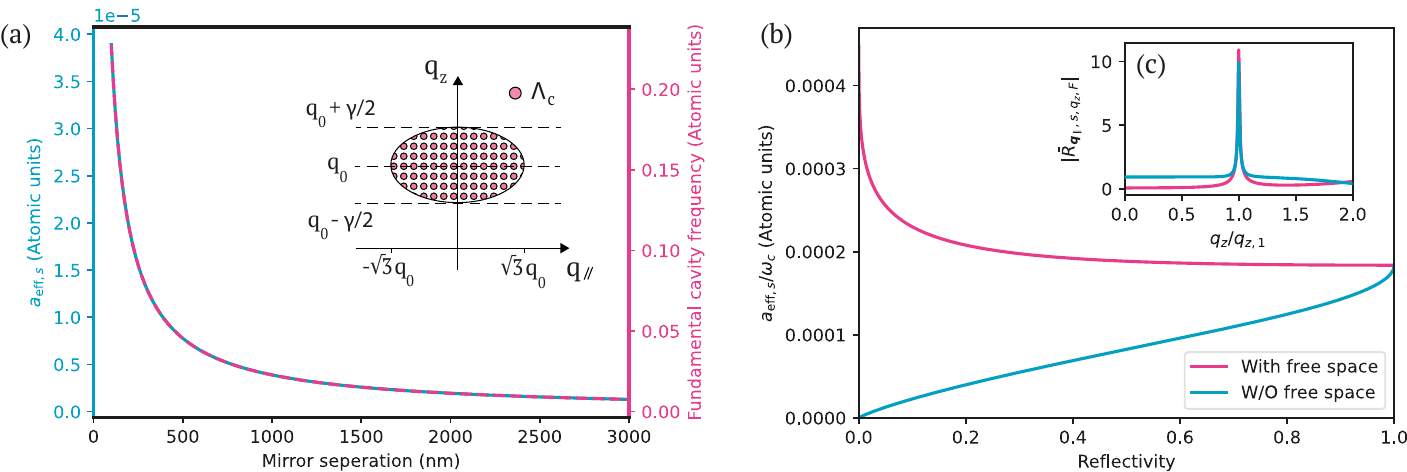}
    \caption{(a) The effective coupling strength (solid light blue) and the fundamental cavity frequency (dashed purple) as a function of mirror seperation, $L_c$ for a fixed reflectivity of 0.99. The inset shows a visualization of the part of $\boldsymbol{q}$-space which contributes to the single effective mode.
    (b) The effective coupling strength $a_{\mathrm{eff},s}/\omega_c$ as a function of the mirror reflectivity for a symmetric Fabry-P\'erot cavity with (purple) and without (light blue) the removal of the free space mode contribution. The ratio $a_{\mathrm{eff},s}/\omega_c$ is independent of the frequency, and it therefore provides a scale-invariant representation of the cavity single effective mode coupling strength in the Fabry-P\'erot cavity.  (c) The cavity mode functions, $|\bar{R}_{\bfq_\parallel,s,q_z,F}|$, with (purple) and without (light blue) the free space contribution.}
    \label{fig:eff_coupling_strenths}
\end{figure*}
\section{The bulk limit of the light-matter coupling}\label{sec:macroscopic_limit_and_length_scales_of_qed}

While already hinted in the previous section, it is worth explicitly discussing what the size scaling of the effective Hamiltonian is, and its consequences on how the bulk limit for a coupled cavity-matter system should be understood. In particular, we see from Eqs.~\eqref{eq:pcut}-\eqref{eq:zcut}  that both in the paramagnetic and the diamagnetic term the dependence on the quantization box volume disappears since $l_\parallel l_z \sim V$. Inserting the expressions for $l_z$ and $l_\parallel$ we find an effective cavity field strength,
\begin{align}\label{eq:g_eff}
    a_{\mathrm{eff},\lambda}= \left(\frac{l_zl_\parallel}{V}\right)^{1/2}A_{\mathrm{eff},\lambda}=\left(\frac{\sqrt{3}}{8L_c^3\mathcal{F}}\right)^{1/2}A_{\mathrm{eff},\lambda}.
\end{align}
We thus arrive at an expression for the coupling strength of the effective single mode in the Fabry-Perot cavity which has no dependence on the volume of the quantization box. The only dependence on the system size left is in number of matter particles which is encoded in the operators $\hat{P}_{\rm el}$ and $\hat{N}_{\rm el}$. These operators scale with the size of the electronic system, and the importance of these terms scale in the same fashion as the $H_{\rm el}$ term when the size of the matter is increased (as explicitly shown in Appendix~\ref{app:MatrixMacroCoupling}). This implies that the effect of the light-matter coupling will not vanish for the matter side of the problem. On the other hand, from the point of view of the light, it becomes clear that if the size of the matter does not grow with the isotropic quantization box, the effect of the matter on the electromagnetic modes might become negligible over the full space and can only be relevant locally, where the cavity is placed. This should not come as surprise since in our setup the matter system is a ``defect'' from the point of view of the free space. \\

Importantly, Eq.~\eqref{eq:g_eff} further tells us that the cavity sets an effective, finite mode volume which is independent of the quantization volume,
\begin{align}
    V_\mathrm{eff} \sim L_c^3\mathcal{F}.
\end{align}
This is noteworthy because it shows that even an open, unbounded cavity like the Fabry-Perot setup naturally sets a finite mode volume. This again emphasizes that the size of the quantization box is an arbitrary length scale which does not affect the nature of the light-matter interaction. 
We can thus ultimately write,
\begin{align}
a_{\mathrm{eff},\lambda} \equiv \frac{1}{\sqrt{V_\mathrm{eff}}}A_{\mathrm{eff},\lambda}.
\end{align}
Where the mode volume is the volume set by the cavity within which to consider the coupled extended light-matter system. The limit where the matter fills the entire in-plane mode-extension is thus the correct "bulk limit" for the coupled cavity-matter system. The practical consequences of such consideration are elaborated in Sec.~\ref{sec:convergence}.

As discussed in Appendix~\ref{app:fpc}, the emergence of a finite mode volume can be qualitatively understood from classical arguments in terms of the longest in-plane length over which the cavity can be expected to mediate a meaningful interaction between two points. Although it is a subject of future investigations, we expect that for a cavity with intrinsic material losses at equilibrium (for which photons cannot actually be dissipated) that the amount of modes counted in the collective canonical transformation should be reduced. Thus intrinsic losses of the cavity should renormalize the above effective volume and lower the effective light-matter coupling strength.

That the effective in-plane area appearing in the coupling of the 2D material and cavity is finite is critical, because QED in general, and in the LWA in particular, is known to exhibit Landau poles. A Landau pole is the largest energy for which a particular version of QED (here PF theory in the LWA) makes sense and it reflects the fact that QED cannot represent all scales simultaneously~\cite{berestetskii1982quantum,PhysRevLett.80.4119,PhysRevLett.93.110405,hainzl2002mass}. The Landau pole can be reached in one of two ways: Either by including arbitrarily high photon frequencies, or by increasing the number of particles in the system. It is therefore critical that the theory itself sets a finite length scale within which the light-matter coupling should be considered.

We finally note that our theory becomes ambiguous when $r\rightarrow 1$. Here $l_z \rightarrow 0$ and $A_{\mathrm{eff},\lambda}$ becomes divergent. 
This problem reflects that we cannot unambiguously connect isotropic (all directions are the same) and homogeneous (all points are equivalent) free space to the anisotropic and inhomogeneous perfectly reflecting Fabry-P\'erot case.
\\

\subsection{Strength of the effective mode parameter}

Following the definitions derived so far, we can explicitly evaluate the value of the $a_{\rm eff}$ parameters that should be used when setting up the effective few-mode Hamiltonian for the light-matter coupled system. The light blue line in Fig~\ref{fig:eff_coupling_strenths}(a) shows the effective coupling strength for the $s$-polarized mode, and the dashed purple line shows the fundamental cavity frequency, as a function of $L_c$ for a fixed mirror reflectivity of 0.99. It can be seen that $a_{\mathrm{eff},\lambda}\propto\omega_c$. This connection is fixed by the geometrically determined resonance condition of the Fabry-P\'erot cavity. It is in agreement with the fact that the enhancement of the electromagnetic density of states in the Fabry-P\'erot cavity is scale invariant if the frequency dependence of the mirror properties, in the range of the resonance, can be neglected~\cite{dutra1996spontaneous}. Because the effective coupling strength scales linearly with $\omega_c$, the ratio $a_{\mathrm{eff}\lambda}/\omega_c$ will be scale invariant. In Fig.~\ref{fig:eff_coupling_strenths}(b), we therefore provide $a_{\mathrm{eff}\lambda}/\omega_c$ as a function of mirror reflectivity. From this plot, one can derive the effective single mode coupling strength for any Fabry-P\'erot cavity setup after specifying its mirror reflectivity and fundamental cavity frequency. \newline
In Fig.~\ref{fig:eff_coupling_strenths}(b), we provide $a_{\mathrm{eff},s}/\omega_c$ both with and without the removal of the free-space contribution, as defined in section~\ref{subsubsec:removefree}. We observe that the scaling of the coupling strength with the mirror reflectivity is qualitatively different in the two cases. We attribute this difference to the fact that as $r\rightarrow 0$, the mode width $\gamma$ diverges. When free space is removed, this is not a problem because $|\bar{R}_{\bfq_\parallel,s,q_z,F}|$ vanishes for $r\rightarrow0$ . However, with the free-space contribution included, $|{R}_{\bfq_\parallel,s,q_z,F}|$ goes to 1, and this means that we are averaging over a wider and wider region of $q_z$. Since all contributions have a finite weight we obtain a divergence. We further note that for $r\rightarrow1$, both cases meet. At this point, the cavity becomes completely sealed off from its surroundings and the remaining free-space contribution becomes negligibly small relative to the cavity-enhanced contribution. The removal of free space in the perfectly reflecting cavity limit therefore interestingly does not significantly affect the effective coupling strength. The above discussion highlights the importance of properly removing the contribution from the free-space modes in general (not perfectly reflecting) cavity setups.\\

\subsection{Note on $k$-point convergence in \textit{ab initio} simulations}
\label{sec:convergence}

The existence of a maximum effective interaction volume/area has practical implications for electronic $k$-point sampling in numerical calculations. The density of $k$-points is a key convergence metric in \textit{ab initio} simulations~\cite{martin2020electronic,enkovaara2010electronic} 
Intuitively, decreasing $\Delta k$, i.e. the $k$-spacing, is equivalent to increasing the number of unit cells in the crystal and hence the extension of the material coupled coherently to the cavity. As the discussion above shows, this should only be done up until the area of the material is equal to the effective mode area. In other words, the $k$-point density can be freely increased as long as $\Delta k\gg \frac{2\pi}{S_\mathrm{eff}^{1/2}}$.

We note that for an optical cavity with resonances in the visible and near infrared regions, $L_c \sim 300\mathrm{nm}-2.5\mathrm{\mu m}$, $\sqrt{\mathcal{F}}\simeq1$ for even modest reflectivities of around $r=0.2$ and this thus tells us that the characteristic cavity coupling length is at least on the order of $L_c$ (see Figure~\ref{fig:app:normalized_effective_spotsize} in Appendix~\ref{app:fpc}). In practice, very few \textit{ab initio} calculations uses such a dense $k$-point sampling before convergence is reached, and $\Delta k\gg \frac{2\pi}{S_\mathrm{eff}^{1/2}}$ is thus naturally fulfilled in most practical simulations.  Despite this, it is worth briefly discussing what happens when $\Delta k \lesssim \frac{2\pi}{S_\mathrm{eff}^{1/2}}$. In this case, the extension of the material becomes bigger than the characteristic cavity length $S_\mathrm{eff}^{1/2}$ which defines the maximal length between two points that can be coupled by cavity photons. The ``extra'' matter should hence not contribute to increasing the coupling, which is not apparent in Eq.~\eqref{eq:lwsinglefinal}. The key observation to recover the appropriate physical behavior is that in Eq.~\eqref{eq:lwsinglefinal} the number of photonic modes $l_\parallel$ has to be redistributed among different electronic k-points. More specifically $l_\parallel \rightarrow l_\parallel\frac{S_\mathrm{eff}^{1/2}\Delta k}{2\pi}$. This redistribution of coupling weight is essentially in place to guarantee that the non-physical infinite-range interaction of the LWA is naturally cut-off by the characteristic cavity coupling length. \\

\section{Conclusion and Discussion}\label{sec:end_sec}

In this work we have shown that even in an extended, unbounded electromagnetic environment, such as a Fabry-P\'erot cavity, the volume over which the interaction of light and matter mediated by the cavity should be considered is finite. This finite interaction volume is large enough to treat matter in its bulk limit. Importantly, we show that this is the limit that should be considered instead of the standard thermodynamic limit of photon-free solid-state physics, when assessing the scaling of the light-matter coupling with system size within QED. While such a limit is naturally described by the explicit inclusion of the multi-mode nature of the electromagnetic field in the coupled light-matter problem, here we have derived an effective few-mode description of the paradigmatic Fabry-P\'erot-2D material cavity-matter system featuring the correct size scaling in the bulk limit and the simplicity of a few-mode theory. Our effective approach is non-perturbative and free from the double counting of the interaction of matter with the vacuum fluctuations of light already present in free space. We have derived our effective schemes according to the following steps:
(i) The photonic modes of the Fabry-P\'erot cavity were quantized in an isotropic space in order to be consistent with the standard quantum-electrodynamical description of free space. (ii) We removed the spurious free-space double counting by identifying and subtracting the contribution of the cavity in the limit of mirrors with zero reflectivity. (iii) From the realistic cavity reflectivity and by employing the long wavelength approximation, we determined the relevant interaction length scales to define the full Hamiltonian of the effective single mode theory. \\
With light-matter coupling remaining finite in the bulk limit it is possible to expect the cavity to have an effect on the the ground state properties of an extended system. Importantly our effective Hamiltonian can be used as a computationally convenient starting point for refined many-body methods. On the one hand, quantum-electrodynamical density-functional theory (QEDFT)~\cite{ruggenthaler2014quantum,ruggenthaler2015ground,Jestadt2019, dreizler2012density, ullrichTDDFT2011}, and its newly developed functionals~\cite{pellegrini2015optimized,flick2018ab,flick2019light,schafer2021making,flick2022simple}, can be augmented with a realistic few-mode description of the quantum electromagnetic environment without increase in its computational cost. On the other hand, within many-body perturbation theory, the effective Hamiltonian introduces a simplified starting point to build an effective perturbation theory, by eliminating the explicit consideration of photon momentum degrees of freedom. It will be subject of future work to show how starting a perturbation theory from the original multi-mode Pauli-Fierz Hamiltonian and following the same approximations motivated by the long wavelength assumption could lead to an equivalent simplified effective perturbation theory. Further effort should then be devoted to delineate the span of correlation functions that can be calculated with such theory.

To conclude, our work represents an important step towards the first principles simulation of realistic cavity-material engineering while providing an intuitive description of the degrees-of-freedom at play in interacting extended light-matter systems.

\begin{acknowledgments}
%
C.S. acknowledges funding from the Horizon Europe research and innovation program of the European Union under the  Marie Sklodowska-Curie grant agreement no. 101065117 and the Swedish Research Council (VR) through Grant No. 2016-06059. The Flatiron Institute is a division of the Simons Foundation. We acknowledge support from the Max Planck-New York City Center for Non-Equilibrium Quantum Phenomena.
This work was supported by the Cluster of Excellence Advanced Imaging of Matter (AIM), Grupos Consolidados (IT1249-19) and SFB925.
This work was supported by the European Research Council (ERC-2024-SyG- 101167294 ; UnMySt
We acknowledge sup-
port from the European Union Marie Sklodowska-Curie
Doctoral Network
SPARKLE grant No. 101169225.
\end{acknowledgments}
\bibliography{reference}

\clearpage

\newpage
\appendix
\section{Expansion of the transverse electromagnetic field in the Fabry-P\'erot cavity}\label{app:fpc}

In this appendix we discuss the expansion of the transverse electromagnetic field in the Fabry-P\'erot cavity~(FPC). 

\begin{figure}[th]
    \centering
    \includegraphics[width=0.75\linewidth]{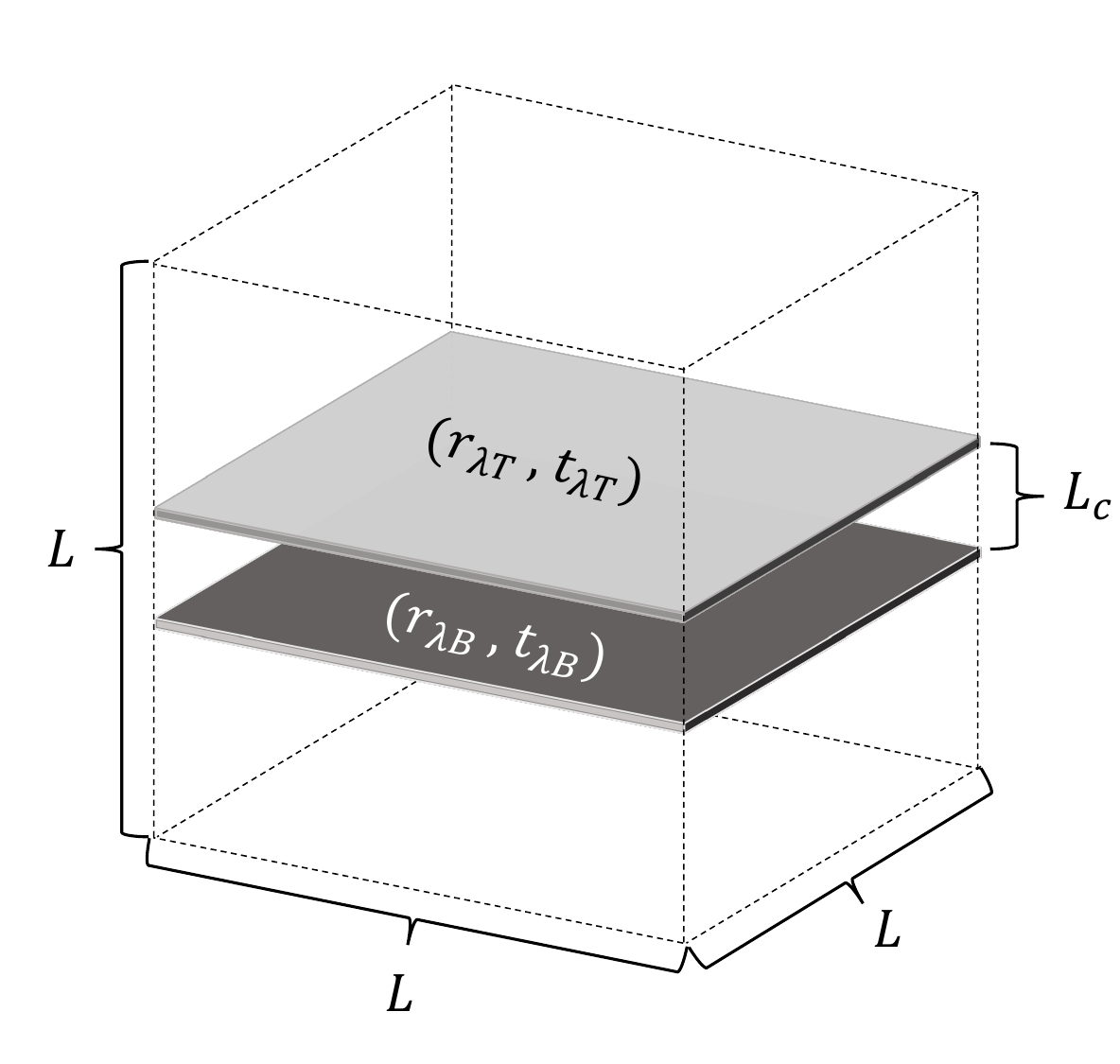}
    \caption{Setup considered for the quantization of the electromagnetic field in the paradigmatic Fabry-Perot cavity}
    \label{fig:app:fpc_quantization_setup}
\end{figure}

To avoid the complications related to gauge invariance discussed in Appendix \ref{app:mismatch}, we follow Ref.~\cite{de1991spontaneous} and consider a cubic box of sides $L$ as our quantization volume, and introduce two thin mirrors $z = \pm L_c/2$. The quantization setting is shown in Fig. \ref{fig:app:fpc_quantization_setup} and it has three important benefits: (i) we can express the mode functions of the vector potential in terms of alterations to the standard plane wave expansion in free space, (ii) we can separate vacuum contributions from reflection contributions, and (iii) we can construct the mode expansion so as to have a direct asymptotic connection with that of free space in the limit of vanishing mirror reflectivity. 

As also mentioned in the main text, in free space the mode functions of the electromagnetic field can be labelled by a wave vector,
\begin{align}
    \bfq = q(\sin\theta\cos\phi, \sin\theta\sin\phi,\cos\theta).
\end{align}
where $\phi\in[0,2\pi]$ and $\theta\in[0,\pi]$. In our model of the FPC, we introduce mirrors which break translational invariance in the out-of-plane direction and couples the two wave vector components,
\begin{align}
    \bfq_\mathrm{T} = q(\sin\theta\cos\phi, \sin\theta\sin\phi,\cos\theta),
\end{align}
and,
\begin{align}
    \bfq_\mathrm{B} = q(\sin\theta\cos\phi, \sin\theta\sin\phi,-\cos\theta).
\end{align}
It is therefore natural to use $\bfq_{\mathrm{T,B}}$ as the basis of wave vectors and restrict $\theta\in[0,\pi/2]$. From the wave vectors and the requirement that $\bfq\cdot\beps_{\bfq_\parallel,\lambda, q_z \alpha} = 0$ we can further construct the polarization vectors,
\begin{align}
    &\beps_{\bfq_\parallel,s, q_z \mathrm{(T,B)}} = (\sin\phi, -\cos\phi, 0)\label{app:eq:polarization:s} \\
    &\beps_{\bfq_\parallel,p, q_z \mathrm{T}} = (\cos\theta\cos\phi, \cos\theta\sin\phi,\sin\theta)\label{app:eq:polarization:pt} \\
    &\beps_{\bfq_\parallel,p, q_z \mathrm{B}} = (\cos\theta\cos\phi, \cos\theta\sin\phi,-\sin\theta) \label{app:eq:polarization:pb}
\end{align}
which naturally correspond to the standard $s$ and $p$ polarizations normally used to describe reflection and transmission in layered media. The coupling between the upwards and downwards travelling waves can be then determined using standard transfer optics methodology~\cite{saleh2019fundamentals}.
Because the cavity mixes modes with $\bfq_\mathrm{T,B}$, the mode functions originally coming from $\bfq_\mathrm{T,B}$ will also have components of the opposite wave vector. These components are respectively,
\begin{align}
    &U_{\bfq_\parallel\lambda, q_z \mathrm{T}}(\bfr) =
    \begin{cases}
            t_{1,\lambda}e^{i\bfq_\mathrm{T}\bfr}/D_\lambda, &         \rm{upwards},\\
            t_{1,\lambda}r_{2,\lambda}e^{i\bfq_\mathrm{B}\bfr +iq_zL_c}/D_\lambda, &         \rm{downwards}.
    \end{cases} \\
    &U_{\bfq_\parallel\lambda, q_z \mathrm{B}}(\bfr) =
    \begin{cases}
            t_{2,\lambda}e^{i\bfq_\mathrm{B}\bfr}/D_\lambda, &         \rm{upwards},\\
            t_{2,\lambda}r_{1,\lambda}e^{i\bfq_\mathrm{T}\bfr +iq_zL_c}/D_\lambda, &         \rm{downwards},
    \end{cases}\\
    & \rm{with,} \nonumber \\
    &\int d\bfr \beps_{\bfq_\parallel,p, q_z \alpha}\cdot\beps_{\bfq_\parallel',\lambda', q_z', \alpha'} U_{\bfq_\parallel\lambda, q_z \alpha}(\bfr)\left(U_{\bfq_\parallel'\lambda', q_z' \alpha'}(\bfr)\right)^* \nonumber \\ &=(2\pi)^3\delta_{\alpha\alpha'}\delta_{\lambda\lambda'}\delta(\bfq - \bfq').
\end{align}
We defined $q_z = q\,\cos\theta$ and $\lambda=s,p$. $r_{n,\lambda}$ and $t_{n,\lambda}$ are the Fresnel reflection and transmission coefficients of the mirrors, and $D_\lambda = 1 - r_{1\lambda}r_{2\lambda}e^{2iq_zL_c}$. Importantly, the wave vector determines which of the polarization functions to use when writing the vector potential. Notice that if we make one of the mirrors perfect s.t. e.g. $t_{2,\lambda} = 0$ ($t_{1,\lambda} = 0$), we loose the contributions from backward (forward) waves.

Next we expand the vector potential in terms of the mode functions. To make the polarizations more explicit, we absorb them into the new vector mode functions and write,
\begin{widetext}
\begin{align}\label{eq:app-fpc:expansion_in_fpc_via_Ms}
    \hat{\bfA}(\bfr) = \sqrt{\frac{1}{V}}\sum_{\bfq_\parallel,\lambda}^{\bfq_{\rm c,\parallel}^{\rm lw}}\sum_{q_z,\alpha}^{\bfq_{\rm c, z}^{\rm lw}} A_{0,\bfq_\parallel q_z}e^{i\bfq_\parallel\cdot\bfr_\parallel}\left(\bfM_{\bfq_\parallel\lambda, q_z\alpha}(z)\hat{a}_{\bfq_\parallel\lambda,q_z\alpha} +\bfM_{-\bfq_\parallel\lambda,q_z\alpha}^*(z)\hat{a}^{\dagger}_{-\bfq_\parallel\lambda,q_z\alpha}\right).
\end{align}
\end{widetext}
where,
\begin{align}
    &\bfM_{\bfq_\parallel\lambda, q_z \mathrm{T}}(z)\nonumber\\
    &= (2\pi)^{3/2}\frac{t_{1,\lambda}e^{iq_zz}\beps_{\bfq_\parallel,\lambda, q_z \mathrm{T}} + t_{1,\lambda}r_{2,\lambda}e^{-iq_z(z-L_c)}\beps_{\bfq_\parallel,\lambda, q_z \mathrm{B}}}{D_\lambda}\\
    &\bfM_{\bfq_\parallel\lambda, q_z \mathrm{B}}(z)\nonumber \\
    &= (2\pi)^{3/2}\frac{t_{2,\lambda}e^{-iq_zz}\beps_{\bfq_\parallel,\lambda, q_z \mathrm{B}} + t_{2,\lambda}r_{1,\lambda}e^{iq_z(z+L_c)}\beps_{\bfq_\parallel,\lambda, q_z \mathrm{T}}}{D_\lambda}
\end{align}
and $A_{0,\bfq_\parallel q_z} =\sqrt{\frac{4\pi}{\omega_{\bfq}}}$ where $\omega_{\bfq_\parallel,q_z} = c\sqrt{q_\parallel^2 + q_z^2}$. Here the operators create photons in the respective modes and obey,
\begin{align}
    \left[\hat{a}_{\bfq_\parallel\lambda,q_z\alpha},\hat{a}^\dagger_{\bfq_\parallel'\lambda',q_z'\alpha'}\right] = \delta_{\alpha\alpha'}\delta_{\lambda\lambda'}\delta_{\bfq_\parallel\bfq_\parallel'}\delta_{q_zq_z'}.
\end{align}

\begin{figure}
    \centering
    \includegraphics[width=0.8\linewidth]{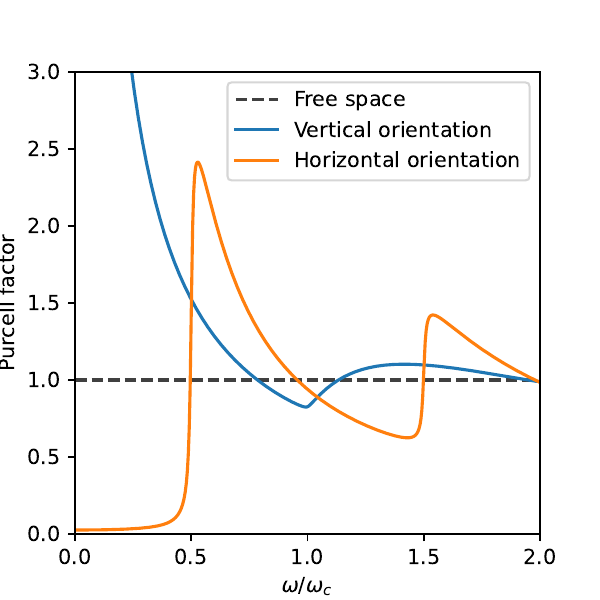}
    \caption{Mode expansion benchmark: Purcell enhancement for point emitters as a function of $\omega/\omega_c$ for a mirror reflectivity of 0.95 as calculated using the expansion of the vector potential presented in Eq.~\ref{eq:app-fpc:expansion_in_fpc_via_Ms}. These agree well with the well known results in the literature~\cite{dutra1996spontaneous,svendsen2023molecules}.}
    \label{fig:app:point_emitter_rates_benchmark}
\end{figure}

To confirm that our expansion behaves as we expect, we have used it to calculate frequency dependent Purcell enhancement for point dipoles with both horizontal- and vertical dipole moments. As shown in Fig.~\ref{fig:app:point_emitter_rates_benchmark}, these agree well with the well known results in the literature~\cite{dutra1996spontaneous,svendsen2023molecules}.

\subsection{The cavity as an effective dressing on the electromagnetic field}
\label{app:dressinglight}
In cases where it can be assumed that the extension of the material in the out-of-plane direction is much smaller than the one of the cavity, we can replace the polarization functions of the cavity with their value at the vertical position of the material. Here we consider the case where the material is placed in the center of the cavity and therefore it is convenient to define $\bar{\bfM}_{\bfq_\parallel\lambda,q_z\alpha}=\bfM_{\bfq_\parallel\lambda,q_z\alpha}(z=0)$. As a next step, we define a new set of operators by absorbing the mode functions into the operator definitions,
\begin{align}
\label{eq:new_operators}
    &\boldsymbol{\hat{b }}_{\bfq_\parallel\lambda q_z\alpha} \equiv \frac{\bar{\bfM}_{\bfq_\parallel\lambda,q_z\alpha}}{|\bar{\bfM}_{\bfq_\parallel\lambda,q_z\alpha}|} \hat{a}_{\bfq_\parallel\lambda q_z\alpha} \\
    &\boldsymbol{\hat{b}}^{\dagger}_{\bfq_\parallel\lambda q_z\alpha} \equiv \frac{\bar{\bfM}^*_{\bfq_\parallel\lambda,q_z\alpha}}{|\bar{\bfM}_{\bfq_\parallel\lambda,q_z\alpha}|} \hat{a}_{\bfq_\parallel\lambda q_z\alpha}^{\dagger} 
\end{align}
which inherit the commutation relations from the $\hat{a}_{\bfq_\parallel\lambda q_z\alpha},\hat{a}_{\bfq_\parallel\lambda q_z\alpha}^\dagger$ operators such that,
\begin{align}
    \left[\boldsymbol{\hat{b}}_{\bfq_\parallel\lambda q_z\alpha}, \boldsymbol{\hat{b}}^{\dagger}_{\bfq_\parallel'\lambda' q_z'\alpha'}\right] = \delta_{\bfq_\parallel,\bfq'_\parallel}\delta_{\lambda,\lambda'}\delta_{q_zq_z'}\delta_{\alpha,\alpha'}
\end{align}
and all other commutators vanish. In terms of these, we can then ultimately expand the vector potential as,
\begin{align}
    \hat{\bfA}(\bfr) = \frac{1}{\sqrt{V}}\sum_{\bfq_\parallel,\lambda}^{\bfq_{\rm c,\parallel}^{\rm lw}}\sum_{q_z,\alpha}^{\bfq_{\rm c, z}^{\rm lw}} A_{c,\bfq_\parallel\lambda q_z\alpha}e^{i\bfq_\parallel\cdot\bfr_\parallel}\left(\boldsymbol{\hat{b}}_{\bfq_\parallel\lambda q_z\alpha} +\boldsymbol{\hat{b}}^{\dagger}_{-\bfq_\parallel\lambda q_z\alpha}\right) 
\end{align}
where we have defined $A_{c,\bfq_\parallel\lambda q_z\alpha} \equiv A_{0,\bfq_\parallel q_z} |\bar{\bfM}_{\bfq_\parallel\lambda,q_z\alpha}|$. When we can restrict the description to a single vertical position, we therefore see that the effect of the cavity is to change the prominence of different wave vectors in the cavity. We can almost think of the mirrors as entering as an effective ``dressing" of the free space modes of the electromagnetic field. Specifically, from the functional dependence of $|\bar{\bfM}_{\bfq_\parallel\lambda,q_z\alpha}|$ on $q_z$, we observe that it enhances $q_z$ which are close to the standing wave conditions in the perfect FPC $q_{c,n} = n \frac{\pi}{L_c}$ for $n\in\mathbb{N}$ and suppress the rest.

\subsection{Splitting in- and out-of-plane parts}
\label{app:splittingpol}
Because we neglect any out-of-plane polarization of the material, it is convenient to rewrite the vector mode functions as,
\begin{align}
    \bar{\bfM}_{\bfq_\parallel\lambda,q_z\alpha} = \bar{\bfM}_{\bfq_\parallel\lambda,q_z\alpha}^\parallel + \bar{\bfM}_{\bfq_\parallel\lambda,q_z\alpha}^z
\end{align}
where $\parallel$($z$) refers to the part of $\bar{\bfM}_{\bfq_\parallel\lambda,q_z\alpha}$ which is parallel(perpendicular) to the mirrors. Under the approximation of vanishing out-of-plane momentum matrix element of the matter, we can thus write,
\begin{align}
    \hat{\bfP}_{\rm el}\cdot \bar{\bfM}_{\bfq_\parallel,s,q_z\alpha} = \hat{\bfP}_{\rm el}\cdot \bar{\bfM}_{\bfq_\parallel,s,q_z\alpha}^\parallel.
\end{align}
By inspection of the polarization vectors in Eqs. \ref{app:eq:polarization:s}-\ref{app:eq:polarization:pb} we can confirm that this leaves the $s$-polarization is unchanged since $\bar{\bfM}_{\bfq_\parallel,s,q_z\alpha} = \bar{\bfM}_{\bfq_\parallel,s,q_z\alpha}^\parallel$, but that we loose the perpendicular component for the $p$-polarization. 

\subsubsection{$s$-polarization}
For the $s$-polarization we trivially pull the polarizations out the $\bfM$ and write,
\begin{align}
    \bar{\bfM}^\parallel_{\bfq_\parallel,s,q_z\alpha} = \beps_{\bfq_\parallel,s, q_z \alpha} \bar{R}_{\bfq_\parallel,s,q_z\alpha}
\end{align}
where we generally define,
\begin{align}
    &\bar{R}_{\bfq_\parallel,\lambda,q_z\mathrm{T}}
    = (2\pi)^{3/2}\frac{t_{1,\lambda}e^{iq_zz} + t_{1,\lambda}r_{2,\lambda}e^{-iq_z(z-d)}}{D_\lambda},\\
    &\bar{R}_{\bfq_\parallel,\lambda,q_z\mathrm{B}}
    = (2\pi)^{3/2}\frac{t_{2,\lambda}e^{-iq_zz} + t_{2,\lambda}r_{1,\lambda}e^{iq_z(z+d)}}{D_\lambda}
\end{align}
Notice that $\bar{R}_{\bfq_\parallel,\lambda,q_z\mathrm{T}} = \bar{R}_{\bfq_\parallel,\lambda,q_z\mathrm{B}}$ for symmetric setups where $(r_{1,\lambda}, t_{1,\lambda}) = (r_{2,\lambda}, t_{2,\lambda})$. Here it is therefore trivial to extract the in-plane polarization and we can define new operators as,
\begin{align}
    &
    \hat{b}_{\bfq_\parallel s q_z\alpha} \equiv \frac{\bar{R}_{\bfq_\parallel,s,q_z\alpha}}{|\bar{R}_{\bfq_\parallel,s,q_z\alpha}|} \hat{a}_{\bfq_\parallel s q_z\alpha} \\
    &\hat{b}^{\dagger}_{\bfq_\parallel s q_z\alpha} \equiv \frac{\bar{R}_{\bfq_\parallel,s,q_z\alpha}^*}{|\bar{R}_{\bfq_\parallel,s,q_z\alpha}|} \hat{a}_{\bfq_\parallel s q_z\alpha}^{\dagger} 
\end{align}
Which allows us to make the substitution like in Eq. \ref{eq:new_operators} and renormalize the vector potential matrix element by $|\bar{R}_{\bfq_\parallel,s,q_z\alpha}|$.

\subsubsection{$p$-polarization}
For the $p$-polarization this is more complicated. However, since,
\begin{align}
    &\mathbf{P}_{\rm el}\cdot\beps_{\bfq_\parallel,p, q_z \mathrm{T}} = \mathbf{P}_{\rm el}\cdot\beps^\parallel_{\bfq_\parallel,p, q_z}\\
    &\mathbf{P}_{\rm el}\cdot\beps_{\bfq_\parallel,\lambda, q_z \mathrm{B}} = \mathbf{P}_{\rm el}\cdot\beps^\parallel_{\bfq_\parallel,p, q_z}
\end{align}
the in-plane part is the same for both the top- and bottom propagating waves,
\begin{align}
    \beps^\parallel_{\bfq_\parallel,p, q_z} = \cos\theta(\cos\phi, \sin\phi,0 ).
\end{align}
We can therefore write,
\begin{align}
    \hat{\bfP}_{\rm el}\cdot \bar{\bfM}_{\bfq_\parallel,p,q_z\alpha} = \cos\theta(\hat{\bfP}_{\rm el}\cdot \beps'_{\bfq_\parallel,p, q_z}) \bar{R}_{\bfq_\parallel,p,q_z\alpha}
\end{align}
where $\bar{R}_{\bfq_\parallel,\lambda,q_z\alpha}$ is defined as above and
\begin{align}
    \tilde{\beps}^{\parallel}_{\bfq_\parallel,p, q_z} = (\cos\phi, \sin\phi,0 ).
\end{align}
Since we can write $\cos\theta = \frac{q_z}{\sqrt{q_z^2 + q^2_\parallel}}$ and because $q_z$ is always positive we can define the operators for the p-polarization as,
\begin{align}
    &
    \hat{b}_{\bfq_\parallel p q_z\alpha} \equiv \frac{\bar{R}_{\bfq_\parallel,p,q_z\alpha}}{|\bar{R}_{\bfq_\parallel,p,q_z\alpha}|} \hat{a}_{\bfq_\parallel p q_z\alpha} \\
    &\hat{b}^{\dagger}_{\bfq_\parallel p q_z\alpha} \equiv \frac{\bar{R}_{\bfq_\parallel,p,q_z\alpha}^*}{|\bar{R}_{\bfq_\parallel,p,q_z\alpha}|} \hat{a}_{\bfq_\parallel p q_z\alpha}^{\dagger} 
\end{align}
and the renormalized vector potential amplitude as, $\frac{q_z}{\sqrt{q_z^2 + q^2_\parallel}}|\bar{R}_{\bfq_\parallel,\lambda,q_z\alpha}|$, the same way it is done in the main text.

\subsection{Estimating the effective spot size in a Fabry-Perot cavity}
\begin{figure}[t!]
    \centering
    \includegraphics[width=0.3\textwidth]{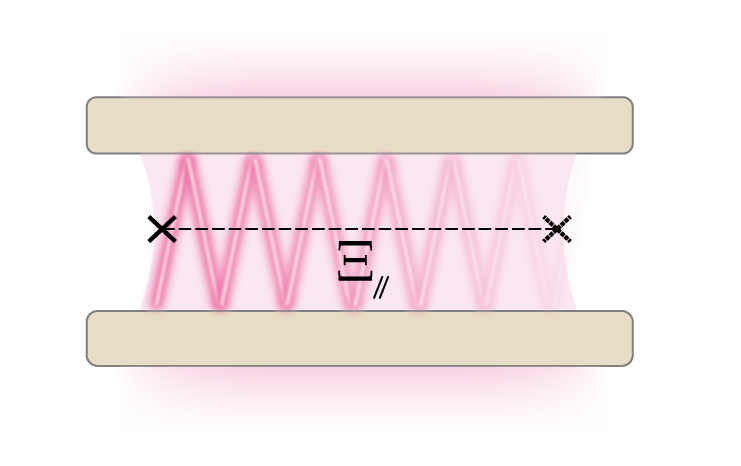}
    \caption{The effective cavity interaction distance is related to the number bounces that light can take inside the cavity before completely leaking out.}
    \label{fig:cavity_bounces}
\end{figure}

In the following we seek an estimate of the effective spot size in the Fabry-Perot cavity. By effective spot size we understand the maximum distance between two points for which it is reasonable to expect the cavity to induce significant correlations. Classically, the distance a given mode would travel inside the cavity is determined by the mirror reflectivity and its angle of incidence. The former determines how many reflections the light can undergo before being damped below some threshold (see Fig.~\ref{fig:cavity_bounces} for an illustration), and the latter determines how far the light will travel between each bounce. 
By combining the two we can therefore get an estimate of the maximum distance between two points for which a given mode should be able to carry interaction.

At an incidence angle of $\theta$, the light will travel the following distance between each reflection on one of the cavity mirrors,
\begin{align}
    \eta = \frac{L_c}{\mathrm{cos}(\theta)}
\end{align}
Between successive reflections, the light therefore covers an in-plane distance of,
\begin{align}
    \eta_\parallel = L_c \mathrm{tan}(\theta)
\end{align}
The reflectivity of the mirrors determine how many reflections the light can undergo before it is attenuated enough such that it is no longer relevant. The total in-plane length that a given mode can resolve together should thus be,
\begin{align}
    \Xi_\parallel = n_r \eta_\parallel
\end{align}
where $n_r$ is the number of allowed reflections.

We can define the number of allowed reflections in our theory in terms of the mirror reflectivity by defining some attenuation cut-off $\epsilon$ and say that the number of bounces is the largest $n_r$ for which,
\begin{align}
    |r|^{n_r} \geq \epsilon
\end{align}
where $r$ is the mirror reflectivity. For any non-unity reflectivity, $n_r$ is finite for any epsilon and there is therefore a largest in-plane region that can be resolved in the cavity. Here will take $\epsilon$ to be $1/e$. The total in-plane distance resolved by the mode is thus approximately,
\begin{align}
    \Xi_\parallel\approx -\eta_\parallel \frac{1}{\ln |r|}
\end{align}
which clearly diverges when $r = 1$. This highlights the competition between the spotsize defined by the cavity properties and the Landau pole breakdown of the long wavelength approximate QED.

\subsection{Effective spot size of the effective single mode}
\begin{figure}
    \centering
    \includegraphics[width=0.9\linewidth]{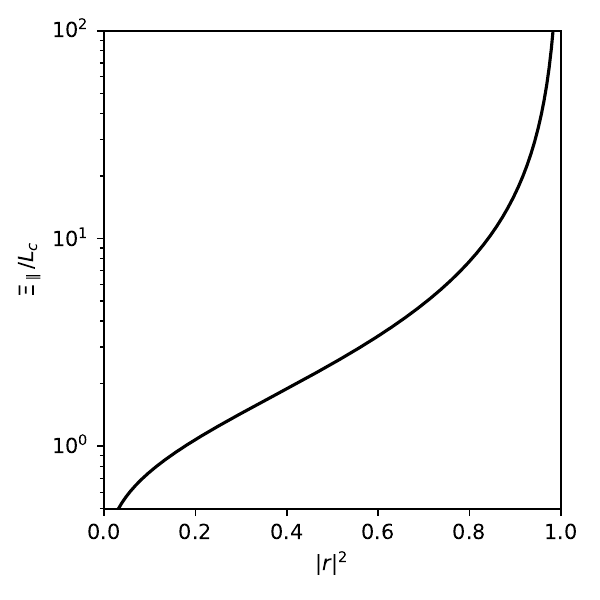}
    \caption{Spot size of the effective single mode normalized by the mirror seperation $L_c$ as a function of mirror reflectivity following Eq.~\ref{eq:app-fpc:effective_spot_size}}
    \label{fig:app:normalized_effective_spotsize}
\end{figure}

In terms of the wave-vector we can write,
\begin{align}
    \mathrm{tan}(\theta) = \frac{q_\parallel}{q_z}
\end{align}
As discussed in the main text, the role of the cavity is to fix $q_z$. The largest $q_\parallel$ that our effective theory contains is $q_\parallel^\mathrm{max} = \sqrt{3}q_{z,1}$ meaning that,
\begin{align}\label{eq:app-fpc:effective_spot_size}
    \Xi_\parallel\approx -\frac{\sqrt{3}L_c}{\ln |r|}.
\end{align}
In Fig.~\ref{fig:app:normalized_effective_spotsize} we show the effective spot-size of the single effective mode as a function of mirror reflectivity. It is clearly observed that it is strongly reflectivity dependent and that it diverges at $|r|^2\rightarrow 1$ as expected. In practice, mirror imperfections will likely make the effective spot-size/effective mode volume smaller than this upper bound.

\section{Incommensurability issues for Crystalline Systems}
\label{app:mismatch}
In the following we discuss the complications which can arise from a mismatch between the symmetry of the Maxwell system associated with the photons and of the periodic crystalline system. The example we give is between the usual cubic symmetry for the free-space electromagnetic field quantization and the periodic boundary condition of a non-cubic crystal group. While this problem is naturally circumvented by the LWA it is of importance when addressing the multi-mode nature of the coupling beyond the LWA. For the free-space electromagnetic field we assume the isotropy of space in agreement with the (special-relativistic) symmetries of the Poincar\'e group. From the local conservation of charge in this cubic quantization volume, i.e., that the local $U(1)$ gauge freedom of the photon field is connected to the phase of the matter wave function~\cite{ruggenthaler2022understanding}, we can deduce the minimal-coupling prescription $\mathbf{\hat{p}} \rightarrow \mathbf{\hat{p}} + \bAh_{\perp}$. Here the (matter) momentum operator and the electromagnetic field necessarily share the same plane waves $
\varphi_{\bfq}(\bfr) = \tfrac{1}{\sqrt{L^3}} e^{\imagi \bfq\cdot\bfr},
$
where $\bfq=\tfrac{2 \pi}{L}\bfn$ with $\bfn \in \mathbb{Z}^3_0$ and $L$ the length of the cubic quantization volume. The main difference between light and matter is that for the Maxwell field the eigenmodes are vectorial, i.e., $-\bnabla^2 \boldsymbol{\varphi}(\bfr) = q^2 \boldsymbol{\varphi}(\bfr)$. The vectorial eigenmodes are then given by  
$\boldsymbol{\varphi}_{\bfq \lambda}(\bfr) = \varphi_{\bfq}(\bfr)\beps(\bfq \lambda)$ with $\beps(\bfq \lambda)$ the two transverse ($\lambda \in \{1,2\}$) and the longitudinal ($\lambda=3$) linear polarization vectors~\cite{greiner1996field}.

If we now consider a non-cubic system, the gauge principle (local conservation of charge) would imply that one also has to use a non-cubic quantization volume for the photonic system. So different directions are different in contrast to the isotropy assumed for free-space fields and we would consider the extended system coupled to a different (not free-space) photon field in contrast to our initial starting point.

A potential way around this issue is to embed the cubic quantization box of the photon field and the non-cubic shape of the crystal in the same $\mathbb{R}^3$ and try to match them in some way. Clearly, since we have now physically different shapes, the local conservation of charges (gauge principle) is not immediately applicable to couple light and matter. However, assuming that the crystal system is contained in the photon quantization volume we can still re-express the (non-matching) eigenmodes of the photon field in the non-cubic volume. For the vectorial eigenfunctions of the free-space photon field this means we have
\begin{align}
    \boldsymbol{\varphi}_{\bfq \lambda}(\bfr) = \sum_{\bfq' \lambda' } \underbrace{\left(\int_{V} \boldsymbol{\phi}_{\bfq' \lambda'}^{*}(\bfr')\cdot \boldsymbol{\varphi}_{\bfq \lambda}(\bfr') d\bfr'\right)}_{=U_{\bfq' \lambda', \bfq \lambda}} \boldsymbol{\phi}_{\bfq' \lambda'}(\bfr)
\end{align}
where $\bfr$ is now restricted to the non-cubic crystal and $\boldsymbol{\phi}_{\bfq' \lambda'}(\bfr)$ are the normalized eigenfunctions of the vectorial Laplacian of the non-cubic crystal with volume $V$. Using the Bloch expansion and restricting $\bfk$ to the first BZ of the non-cubic system we can rewrite further $\boldsymbol{\varphi}_{\bfq \lambda}(\bfr) = \sum_{\lambda' \bfk \bfG} U_{\bfk+\bfG \lambda', \bfq \lambda} \boldsymbol{\phi}_{\bfk+\bfG \lambda'}(\bfr)$, where $\bfG$ are all possible reciprocal lattice vectors. If we, in a next step, define
\begin{align}
     \hat{a}_{\bfk+\bfG, \lambda'}(\bfr) = \sqrt{V} \sum_{\bfq \lambda} U_{\bfk+\bfG \lambda', \bfq \lambda} \boldsymbol{\varphi}_{\bfq, \lambda}(\bfr) \hat{a}_{\bfq,\lambda} 
\end{align}
we can re-express
\be
\label{eq:vectorpotential_incommensurate}
    \bAh_{\perp}(\bfr)^{\rm free} \rightarrow \sqrt{\frac{(2\pi)^3}{V}}\sum_{\bfk \bfG \lambda'}  \left(\hat{a}_{\bfk+\bfG,\lambda'}(\bfr)  + h.c. \right),
\ee

Note that because of the incommensurability, even small $\bfq$ components of the vector potential will be projected onto very large crystal momenta.
In this notation we again see that we have all possible periodicities in the photon field and hence the Hamiltonian breaks the usual simple matter translational invariance.

In a similar way we can also express the longitudinal interaction kernel in terms of the non-cubic unit cell.

We note, however, that since we have a mismatch between the light and the matter sector violating the basic gauge principle, it remains unclear whether such a procedure leads to sound results (for a detailed discussion see Ref.~\cite{ruggenthaler2022understanding}). 
While this discussion remains of crucial importance when explicitly accounting the momentum transfer from photons to matter, the long wavelength approximation resolves the issues, because it is no longer required to have a one-to-one matching of photon and crystal momenta.

\section{Derivation of the Pauli-Fierz Hamiltonian Bloch representation}
\label{app:PFBloch}
In this appendix we derive the Pauli-Fierz Hamiltonian making use of the system translational invariance which allows us to express electronic field operators in Bloch's function. Let us first consider a many-body electronic system 
\begin{equation}
\begin{split}
\hat{H}^{\rm{el}} = &\sum_{\sigma}\int_{V_{\rm M}}{ d\bfr \hat{\Psi}^{\dagger}(\bfr \sigma)\left[-\frac{\nabla^2}{2}-\phi(\bfr)\right]\hat{\Psi}(\bfr \sigma)} + \\
& \sum_{\sigma\sigma'}\int_{V_{\rm M}} d\bfr d\bfr' \hat{\Psi}^{\dagger}(\bfr \sigma)\hat{\Psi}^{\dagger}(\bfr' \sigma')v(\bfr,\bfr')\hat{\Psi}(\bfr' \sigma')\hat{\Psi}(\bfr \sigma),
\end{split}
\end{equation}
where $V_{\rm M}$ is the volume of the matter (such as $V_{\rm M}=N_{\rm c}\Omega$ and $\Omega$ the unit cell volume), $\hat{\Psi}^{\dagger}(\bfr \sigma),\hat{\Psi}^{\dagger}(\bfr \sigma)$ the creation and annihilation operators of electrons at point $\bfr$ with spin $\sigma$ and $\phi(\bfr)$ the single-particle potential due to the nuclei of the crystal. In the case of a periodic electronic system, as in a crystal, the creation and annihilation operators can be conveniently expressed in the basis of Bloch wave functions (here we consider a system periodic in the in two dimensions as in the main text)
\begin{equation}
\hat{\psi}(\bfr \sigma) = \frac{1}{\sqrt{S_{\rm M}}}\sum_{i\bfk_\parallel \sigma}^{\rm BZ}e^{i\bfk_\parallel\cdot(\tilde{\bfr}+\bfR_\parallel)}u_{i\bfk_\parallel}(\tilde{\bfr} \sigma)\hat{c}_{i\bfk_\parallel \sigma},
\end{equation}
with $u_{i\bfk_\parallel}(\tilde{\bfr} \sigma)$ the periodic part of the Bloch wave function (which satisfies $u_{i\bfk_\parallel}((\bfr=\tilde{\bfr}+\bfR_\parallel)\sigma)=u_{i\bfk_\parallel}(\tilde{\bfr} \sigma)$ with $\bfR_\parallel$ a generic real space lattice vector) and $\hat{c}_{i\bfk_\parallel\sigma}$ the annihilation operator of an electron in a Bloch function with index $i\bfk_\parallel$.

We now consider the coupling to an electromagnetic field described by the vector potential in Eq.~\ref{eq:AfieldFPC_all}. The Hamiltonian associated with the free-photon system can be written as:
\be
\hat{H}_{\rm{EM}}=\sum_{\bfq_\parallel \lambda}\sum_{q_z \alpha}\left(\omega_{\bfq_\parallel q_z}+\frac{1}{2}\right)\hat{a}^{\dagger}_{\bfq_\parallel \lambda, q_z \alpha}\hat{a}_{\bfq_\parallel \lambda, q_z \alpha}.
\ee
In order to couple light with matter we proceed with the minimal coupling substitution, $-i\nabla\rightarrow -i\nabla-\bAh(\bfr)$, to obtain the non-relativistic Pauli-Fierz Hamiltonian in Coulomb gauge (ignoring the Stern-Gerlach term~\cite{spohn2004dynamics})
\begin{widetext}
\be
\label{eq:pfmultimodesapp}
\begin{split}
\hat{H}_{\rm{PF}}=&\sum_{\bfq_\parallel \lambda}\sum_{q_z \alpha}\left(\omega_{\bfq_\parallel q_z}+\frac{1}{2}\right)\hat{a}^{\dagger}_{\bfq_\parallel \lambda, q_z \alpha}\hat{a}_{\bfq_\parallel \lambda, q_z \alpha} + \frac{1}{S_{\rm M}}\sum_n\sum_{ij\sigma}\sum_{\bfk_\parallel\bfk_\parallel'}\int_{\Omega} d\tilde{\bfr} e^{-\imagi\bfk_\parallel'\cdot(\tilde{\bfr}+\bfR_n)}u^*_{i\bfk_\parallel'}(\tilde{\bfr} \sigma)\hat{c}^{\dagger}_{i\bfk_\parallel'\sigma}\\
&\left[\frac{1}{2}\left(-\imagi\nabla+\frac{1}{\sqrt{V}}\sum_{\bfq_\parallel,\lambda}\sum_{q_z, \alpha} A_{0\bfq}e^{i\bfq_\parallel\cdot\mathbf{r}}(\bfM_{\bfq_\parallel\lambda, q_z\alpha}(z)\hat{a}_{\bfq_\parallel\lambda, q_z\alpha} +\bfM_{-\bfq_\parallel\lambda,q_z\alpha}^*(z)\hat{a}^{\dagger}_{-\bfq_\parallel\lambda, q_z\alpha})\right)^2- \phi(\tilde{\bfr})\right]\\
&e^{\imagi\bfk_\parallel\cdot(\tilde{\bfr}+\bfR_n)}u_{j\bfk_\parallel}(\tilde{\bfr} \sigma)\hat{c}_{j\bfk_\parallel\sigma} +  \sum_{\sigma\sigma'}\sum_{ijmn}\sum_{\bfk_\parallel\bfk_\parallel'}v_{ijmn, \bfk_\parallel\bfk_\parallel'}\hat{c}^{\dagger}_{i\bfk_\parallel\sigma}\hat{c}^{\dagger}_{j\bfk_\parallel'\sigma'}\hat{c}_{m\bfk_\parallel'\sigma'}\hat{c}_{n\bfk_\parallel\sigma}\\
=&\sum_{\bfq_\parallel \lambda}\sum_{q_z \alpha}\left(\omega_{\bfq_\parallel q_z}+\frac{1}{2}\right)\hat{a}^{\dagger}_{\bfq_\parallel \lambda, q_z \alpha}\hat{a}_{\bfq_\parallel \lambda, q_z \alpha} + \sum_{ij \sigma}\sum_{\bfk_\parallel} h^{\rm{el}}_{ij\sigma\bfk_\parallel}\hat{c}_{i\bfk_\parallel \sigma}^{\dagger}\hat{c}_{j \bfk_\parallel\sigma} + \sum_{\sigma\sigma'}\sum_{ijmn}\sum_{\bfk_\parallel\bfk_\parallel'}v_{ijmn, \bfk_\parallel\bfk_\parallel'}\hat{c}^{\dagger}_{i\bfk_\parallel\sigma}\hat{c}^{\dagger}_{j\bfk_\parallel'\sigma'}\hat{c}_{m\bfk_\parallel'\sigma'}\hat{c}_{n\bfk_\parallel\sigma}+ \\
& \frac{1}{\sqrt{V}}\sum_{ij\sigma\bfk_{\parallel}}\sum_{\bfq_\parallel,\lambda}\sum_{q_z, \alpha}\int_{\Omega_{z}}dz~ \hat{c}_{i\bfk_\parallel+ \bfq_\parallel\sigma}^\dagger\hat{c}_{j\bfk_\parallel\sigma} A_{0\bfq_\parallel q_z}\bfp_{ij\sigma\bfk_{\parallel}\bfq_{\parallel} q_z}(z)\cdot[\bfM_{\bfq_\parallel \lambda, q_z \alpha}(z)\hat{a}_{\bfq_\parallel\lambda,q_z\alpha} +\bfM_{-\bfq_\parallel\lambda,q_z\alpha}^*(z)\hat{a}^{\dagger}_{-\bfq_\parallel\lambda,q_z\alpha}]+\\
&\frac{1}{2V}\sum_{ij\sigma\bfk_\parallel}\sum_{\bfq_\parallel\bfq_\parallel',\lambda\lambda'}\sum_{q_z q_z',\alpha\alpha'}\int_{\Omega_{z}}dz~\hat{c}_{i\bfk_\parallel+\bfq_\parallel+\bfq_\parallel'\sigma}^\dagger\hat{c}_{j\bfk_\parallel\sigma} A_{0\bfq_\parallel q_z}A_{0\bfq_\parallel'q_z'}s_{ij\sigma\bfk_\parallel\bfq_\parallel\bfq_\parallel'}(z)\\
&[\bfM_{\bfq_\parallel\lambda,q_z\alpha}(z)\hat{a}_{\bfq_\parallel\lambda q_z\alpha} +\bfM_{-\bfq_\parallel\lambda,q_z\alpha}^*(z)\hat{a}^{\dagger}_{-\bfq_\parallel\lambda,q_z\alpha}]\cdot[\bfM_{\bfq_\parallel'\lambda',q_z'\alpha'}(z)\hat{a}_{\bfq_\parallel'\lambda',q_z'\alpha'} +\bfM_{-\bfq_\parallel'\lambda',q_z'\alpha'}^*(z)\hat{a}^{\dagger}_{-\bfq_\parallel'\lambda',q_z'\alpha'}],
\end{split}
\ee
\end{widetext}
where in the first line we have explicitly substituted the expressions for the electronic field operators and the vector potential and we have divided the integral over space in a sum over unit cells and integral over the unit cell volume $\Omega$. In the second step we have then defined the following expressions
\be
\begin{split}
&h_{ij\sigma\bfk_\parallel}^{\rm{el}} \equiv \int_{\Omega} d\tilde{\bfr}~ u^*_{i\bfk_\parallel}(\tilde{\bfr}\sigma)\left[\frac{(-\imagi\tilde{\nabla}+\bfk_\parallel)^2}{2}-\phi(\tilde{\bfr})\right]u_{j\bfk_\parallel}(\tilde{\bfr}\sigma)\\
&\boldsymbol{p}_{ij\sigma\bfk_\parallel\bfq_\parallel}(z) \equiv \int_{\Omega_\parallel} d\tilde{\bfr}_\parallel~ u^*_{i\bfk_\parallel+\bfq_\parallel}(\tilde{\bfr}_\parallel \sigma)\left(-\imagi\tilde{\nabla}+\bfk_\parallel\right)u_{j\bfk_\parallel}(\tilde{\bfr}_\parallel\sigma)\\
&s_{ij \sigma\bfk_\parallel\bfq_\parallel\bfq_\parallel'}(z) \equiv \int_{\Omega_\parallel} ~d\tilde{\bfr}_\parallel u^*_{i\bfk_\parallel+\bfq_\parallel+\bfq_\parallel'}(\tilde{\bfr}_\parallel \sigma)u_{j\bfk_\parallel}(\tilde{\bfr}_\parallel\sigma)\\
\end{split}
\ee
where $\Omega_\parallel$ indicates an integration only in the in-plane dimensions of the unit cell. In order to get to the result in the second line of Eq.\ref{eq:pfmultimodesapp} we have used the relation $\delta_{\bfk_\parallel\bfk'_\parallel}=\frac{1}{S_{\rm M}}\sum_n^{N{\rm c}}e^{i(\bfk_\parallel-\bfk'_\parallel)\cdot\bfR_{n}}$. Notice that in the main manuscript the second and third term in the equation above appear as $\hat{H}_{\rm el}$.

\section{Single Effective Mode Hamiltonian. The steps of the derivation}
\label{app:singleeffective}
Here we provide the details for the derivation of the single effective  mode Hamiltonian which have been omitted in the main text. The derivation can be rationalized in three steps.
\subsection{2D thickness limit}
The first assumption is to consider the thickness $d$ of the embedded material to be much smaller than the distance between the mirrors $L_{\rm c}$ so that the function $\bfM_{\bfq_\parallel\lambda,q_z\alpha}(z)$ is essentially a constant where $\boldsymbol{p}_{ij\sigma\bfk\bfq}(z)$ is non-zero and hence can be directly replaced with its value at $z=0$ (defined as $\bar{\bfM}_{\bfq_\parallel\lambda,q_z\alpha}$ in the main text). This assumption allows reduces the integral in the z-direction in Eq.\ref{eq:pfmultimodesapp} to an integral over $\boldsymbol{p}_{ij\sigma\bfk\bfq}(z)$ for the paramagnetic term and over $s_{ij \sigma\bfk_\parallel\bfq_\parallel\bfq_\parallel'}(z)$ for the diamagnetic term. Using the notation of the main text, these are defined as
\be
\begin{split}
\bar{\boldsymbol{p}}_{ij\sigma\bfk_\parallel\bfq_\parallel} &\equiv \int_{\Omega_{z}} dz \boldsymbol{p}_{ij\sigma\bfk_\parallel\bfq_\parallel}(z) \\
\bar{s}_{ij \sigma\bfk_\parallel\bfq_\parallel\bfq_\parallel'} &\equiv  \int_{\Omega_{z}} dz s_{ij \sigma\bfk_\parallel\bfq_\parallel\bfq_\parallel'}(z)
\end{split}
\ee

\subsection{Long-wavelength Approximation}
The next step is to formalize the long-wavelength approximation. Physically, the validity of the long-wavelength approximation. (LWA) requires that the momentum transfer from the photons to the matter is negligible. For the Fabry-Perot cavity discussed in the main text the cut-offs for the photon momenta make the LWA a rather safe assumption for most crystals. In formulas it implies that for any photon momentum we can approximate $\bar{\boldsymbol{p}}_{ij\sigma\bfk_\parallel\bfq_\parallel}\simeq\bar{\boldsymbol{p}}_{ij\sigma\bfk_\parallel\boldsymbol{0}}$ and $\bar{s}_{ij \sigma\bfk_\parallel\bfq_\parallel\bfq_\parallel'}\simeq s_{ij\sigma \bfk_\parallel \boldsymbol{0}\boldsymbol{0}}$. For the latter we can then note that $s_{ij\sigma \bfk_\parallel \boldsymbol{0}\boldsymbol{0}} = \delta_{ij}$ due to the orthonormality of the $u_{i\bfk_\parallel}(\tilde{\bfr}\sigma)$ within the unit cell.
With this first two steps and the light-dressing transformation in App.~\ref{app:dressinglight} we arrive at the long-wavelength Hamiltonian in Eq.~\ref{eq:hamlw}.

\subsection{Collective Canonical transformation}
In this step we proceed with grouping the photon modes in their momentum component and we do so by first approximating all the mode coefficients to $A_{{\rm eff}, \lambda}$ and then define the total displacement operator $\hat{Q}_{{\rm eff}, \lambda}$ in Eq.~\ref{eq:totdispl} as discussed in the main text. Defining $\hat{Q}_{{\rm eff}, \lambda}$ is yet not enough to perform a canonical transformation. 

Indeed one needs to define the relative displacement operators for the $N_{\rm ph}-1$ degrees of freedom left. Here we report the relevant definitions, where for simplicity we do not distinguish between parallel and z-direction (without consequences on the results),
\be
\begin{split}
\hat{Q}_{{\rm eff}, \lambda} &\equiv \frac{1}{\sqrt{2}}\sum_{\bfq}^{\bfq_{\rm c}^{\rm lw}}\left(\hat{b}_{\bfq \lambda}^{\dagger}+\hat{b}_{-\bfq \lambda}\right) = \sum_{\bfq}^{\bfq_{\rm c}^{\rm lw}} \hat{x}_{\bfq \lambda} \\
\hat{\gamma}_{\bfq \lambda} &= \frac{1}{\sqrt{2}}\left(\hat{b}_{\bfq \lambda}^{\dagger}+\hat{b}_{-\bfq \lambda}\right) - \frac{1}{l}\hat{Q}_{{\rm eff},\lambda} = \hat{x}_{\bfq \lambda} - \frac{1}{l}\hat{Q}_{{\rm eff},\lambda}
\end{split}
\ee
where $\hat{Q}_{{\rm eff},\lambda}$ is, once again, the collective displacement of the modes up to $\bfq_{\rm c}^{\rm{lw}}$ with polarization $\lambda$ , $\hat{\gamma}_{\bfq}$ the relative deviation of the displacement of a given mode $\bfq$ with respect to the average mode displacement and $l=l_\parallel l_z$ the number of modes with polarization $\lambda$ up to that mode. The latter obeys the equality $\sum_{\bfq} \hat{\gamma}_{\bfq \lambda}=0$ by construction. The corresponding conjugate operators are given by
\be
\begin{split}
\hat{P}_{{\rm ph}, \lambda} &\equiv -\imagi \frac{\partial}{\partial \hat{Q}} = \sum_{\bfq}^{\bfq_{\rm c}^{\rm lw}} -\imagi \frac{\partial}{\partial \hat{x}_{\bfq \lambda}} \\
-\imagi \frac{\partial}{\partial \hat{x}_{\bfq\lambda}} &\equiv  -\imagi\left[\frac{\partial\hat{Q}_{{\rm eff},\lambda}}{\partial\hat{x}_{\bfq\lambda}}\frac{\partial}{\partial\hat{Q}_{{\rm eff},\lambda}}+\sum_{\bfq'}^{\bfq_{\rm c}^{\rm lw}}\frac{\partial\hat{\gamma}_{\bfq' \lambda}}{\partial\hat{x}_{\bfq \lambda}}\frac{\partial}{\partial\hat{\gamma}_{\bfq' \lambda}}\right]\\
&= \hat{P}_{{\rm ph}, \lambda} -\imagi \sum_{\bfq'}\left[\delta_{\bfq\bfq'}-\frac{1}{l}\frac{\partial\hat{Q}_{{\rm eff},\lambda}}{\partial\hat{x}_{\bfq'\lambda}}\right]\frac{\partial}{\partial\hat{\gamma}_{\bfq' \lambda}}\\
&= \hat{P}_{{\rm ph},\lambda} - \imagi \frac{\partial}{\partial\hat{\gamma}_{\bfq \lambda}}=\hat{P}_{{\rm ph},\lambda} + \hat{p}_{{\rm ph}, \bfq \lambda},
\end{split}
\ee
where we have used the fact that $\sum_{\bfq} \frac{\partial}{\partial\hat{\gamma}_{\bfq \lambda}}=0$ and defined $\hat{p}_{{\rm ph}, \bfq \lambda}=- i \frac{\partial}{\partial\hat{\gamma}_{\bfq \lambda}}$.
With the above operators one can rewrite the free photonic part and the diamagnetic part of the Pauli-Fierz Hamiltonian. In order not to explicitly deal with the diamagnetic term we can simply perform a standard Bogoliubov transformation and dress the free-photon frequency as $\tilde{\omega}_{\rm c} = \sqrt{\omega_{\rm c}^2+\omega_{\rm dia}^2}$. Hence the diamagnetically dressed free-photon Hamiltonian transforms as,
\be
\label{eq:transformation_freedressed}
\begin{split}
\hat{H}_{\rm ph} =& \left(\tilde{\omega}_{\rm c}+\frac{1}{2}\right)\sum_{\bfq \lambda}\hat{b}^{\dagger}_{\bfq \lambda}\hat{b}_{\bfq \lambda} \\
=& \frac{1}{2}\sum_{\lambda}\left(l(\hat{P}_{{\rm ph},\lambda})^2+\tilde{\omega}_{\rm c}^2\frac{(\hat{Q}_{{\rm eff}, \lambda})^2}{l}\right) +\\
&\frac{1}{2}\sum_{\bfq \lambda}\left((\hat{p}^{{\rm ph}, \gamma\bfq})^2+\tilde{\omega}_{\rm c}^2\hat{\gamma}_{\bfq \lambda}^2\right),
\end{split}
\ee
where the last term has been defined as $\hat{H}_{{\rm el, rel}}$ in the main text.
Finally, by defining $\hat{Q}_{{\rm eff}, \lambda}=\sqrt{\frac{l}{2\omega_{\rm c}}}\left(\hat{B}^{\dagger}_{{\rm eff}, \lambda}+\hat{B}_{{\rm eff}, \lambda}\right)$ and $\hat{P}_{{\rm ph}, \lambda}=i\sqrt{\frac{\omega_{\rm c}}{2l}}\left(\hat{B}^{\dagger}_{{\rm eff},\lambda}-\hat{B}_{{\rm eff}, \lambda}\right)$ together with the above listed transformation, and pulling out the diamagnetic contribution in the first term in second step of Eq.~\ref{eq:transformation_freedressed}, we recover Eq.~\eqref{eq:lwsinglefinal}.

\section{Matrix visualization of the macroscopic coupling}
\label{app:MatrixMacroCoupling}
In order to visualize the origin of the macroscopic coupling let us consider $N$ single free electron-hole excitations all coupled via the effective photon mode of the cavity. In a basis set of the type $\{\ket{\Psi_{\rm GS}}, \ket{\Psi_1^{\rm exc}}, \cdots, \ket{\Psi_N^{\rm exc}}\}\otimes\{\ket{0},\ket{1}\}$, i.e. a product state made of electronic and photonic states respectively, the Pauli-Fierz Hamiltonian in Eq.~\eqref{eq:lwsinglefinal} would read
\be
H_{\rm PF} = \begin{bmatrix}
0      & 0      & \cdots &    0   & 0 & C_{\rm eff} & \cdots & C_{\rm eff} \\
0      & \Delta &        &        & C_{\rm eff}^* & 0 &  &  \\
\vdots &        & \ddots &        &  \vdots &  & \ddots &  \\
0       &        &        & \Delta & C^*_{\rm eff} &  &  & 0 \\
0      & C_{\rm eff} & \cdots & C_{\rm eff} & \omega_c & 0 & \cdots & 0 \\
C_{\rm eff}^*        &   0     &        &        &  0        & \Delta + \omega_{\rm c} & &  \\
\vdots                      &        &   \ddots     &        &     \vdots      &                   & \ddots  & \\
C_{\rm eff}^* &  &  & 0  & 0 &  &  & \Delta + \omega_{\rm c}\\
\end{bmatrix},
\ee
for the sake of argument we have assumed that all excitation have the same energy $\Delta$, that they are uncoupled among themselves and that their coupling strength to the ground-state is the same and equal to $C_{\rm eff}\equiv \left(\frac{l_\parallel l_z}{V}\right)^{1/2} A_{\rm eff} \hat{P}_{\rm el}\cdot\bm{\epsilon}_\parallel$. The matrix above can be reshuffled in two blocks: one consisting of matter excitations with no photon, $\{|\Psi_1^{\rm exc}\rangle\, \cdots, |\Psi_N^{\rm exc}\rangle\}\otimes|0\rangle$, coupled to the groundstate plus one photon state, $|\Psi_{\rm GS}\rangle\otimes|1\rangle$, and the other made of groundstate with no photon, $|\Psi_{\rm GS}\rangle\otimes|0\rangle$, coupled to the excited states plus one photon, $\{|\Psi_1^{\rm exc}\rangle\, \cdots, |\Psi_N^{\rm exc}\rangle\}\otimes|1\rangle$. 
We shall now focus on the latter
\be
H_{\rm PF}^{\rm(off-res)} = \begin{bmatrix}
0                   &  C_{\rm eff}       & \cdots   & C_{\rm eff} \\
C_{\rm eff}^*&  \Delta + \omega_{\rm c}  &          &                    \\
\vdots              &                           &  \ddots  &                    \\
C_{\rm eff}^*&                           &          & \Delta + \omega_{\rm c}\\
\end{bmatrix}.
\ee
This matrix representation clearly shows that the groundstate with no phootons can only be coupled off-resonantly to the excited states through the one-photon state. A further reason why this representation is convenient is that it can be easily shown that its associated eigenproblem $H_{\rm PF}^{\rm {off-res}}\Phi = E\Phi$ can be directly mapped onto a 2x2 problem as
\be
H_{\rm PF}^{\rm(off-res)} = \begin{bmatrix}
0   &  NC_{\rm eff}       \\
NC_{\rm eff}^* &  \Delta + \omega_{\rm c} 
\end{bmatrix}\begin{bmatrix}
\Phi_{\rm GS} \\
\Phi_{\rm exc} \\
\end{bmatrix} = E \begin{bmatrix}
\Phi_{\rm GS} \\
\Phi_{\rm exc} \\
\end{bmatrix}
\ee
which in other words means that the change in the groundstate energy, because of the coupling to the N-equivalent excitations will be of order $N$ as expected for an extensive scaling of the electronic Hamiltonian with respect to the number of unit cells in the crystal.

\section{Note on the Grand-Canonical Ensemble in Non-Relativistic QED}
\label{app:grandcanonical}
The use of annihilation and creation operators also for the matter subsystem comes along with some subtle issues that need to be addressed. Firstly, since the matter system is particle-number conserving, the photon field couples to only an individual $N_e$-particle sector and the coupled light-matter Hilbert space is $\mathcal{H}_{N_e} \otimes \mathcal{F}_{+}$, where $\mathcal{H}_{N_e}$ is the fermionic $N_e$-particle Hilbert space and $\mathcal{F}_{+}$ is the bosonic Fock space of the Pauli-Fierz Hamiltonian~\cite{spohn2004dynamics}. If we now promote the matter system to operate on a fermionic Fock space, we need to make sure that we only always restrict to a fixed particle-number sector in all calculations, because we formally work with the (unphysical) coupled Hilbert space $\mathcal{F}_{-}\otimes \mathcal{F}_{+}$. This coupled Hilbert space contains unphysical states, such as wave functions with different particle numbers that couple to the same photonic field. So, in contrast to the statistical interpretation of grand-canonical states in non-relativistic quantum mechanics, where the different particle sectors are statistically independent, the photon field can "feel" the statistical mixtures and unphysically connect the different particle-number sectors. One needs to contrast this to relativistic QED, where the particle number is not conserved (electron-positron pair creation/annihilation) and it is only the total charge that is conserved~\cite{greiner1996field}.

\section{Modification of the Longitudinal Coulomb Interaction}
\label{app:longitudinal}

If we consider how the usual Coulomb interaction arises, and we take into account that a photonic structure changes the modes of the electromagnetic field, we expect to also find a modified longitudinal (Coulomb) interaction between the charged particles of our crystal structure. It is sometimes argued that this would be the major contribution on cavity-induced changes~\cite{de2018cavity,schuler2020vacua}, however we expect such an effect to play a role at smaller length scales (i.e. closer to the mirrors of the cavity) than the ones involved in the case transversal photons. In Coulomb gauge the (for simplicity perfect) cavity has longitudinal eigenmodes $\bfA_{n,\parallel}(\bfr)$ with eigenvalues $q_{n,\parallel}^2$ and $n \in \mathbb{N}$. In order to follow the usual quantization procedure of the light field~\cite{greiner1996field} we then need that the vector and scalar Laplacian share the same spatial eigenfunctions such that $\bfA_{n,\parallel}(\bfr) = \beps(\bfq, \parallel)a_{\bfq}(\bfr)$. Note that this is yet a different consistency that arises between the light and matter sector (compare to App.~\ref{app:mismatch}). In this way we can generate any longitudinal field that is consistent with the Gauss' law and we find the longitudinal interaction as
\begin{align}
    v'(\bfr,\bfr') = \sum_{n} \frac{1}{q_{n,\parallel}^2}a_{\bfq}(\bfr)a_{\bfq}^{*}(\bfr'). 
\end{align}
If we, on the other hand, use more complex models of cavities, e.g., as a linear isotropic dielectric medium~\cite{Glauber1991dielectric} or with the help of macroscopic QED~\cite{buhmann2007dispersion}, a direct formulation of the changed longitudinal interaction is less straightforward.
Finally we stress that screening of the longitudinal interaction can be already captured by standard ab initio many-body approaches~\cite{latini2015excitons} and if needed can be added to the ab initio description on top of the effect from the transversal fields analyzed in this work.

\end{document}